\documentclass[12pt, preprint]{aastex}

\usepackage{natbib}

\newcommand{\lp}{\ell_P}

\newcommand{\nucii}{\nu_{158}}
\newcommand{\Rab}{R_{AB}}
\newcommand{\Snu}{S_{\nu}}
\newcommand{\Lnu}{L_{\nu}}
\newcommand{\Dnu}{\Delta \nu}
\newcommand{\Lcii}{L_{\rm CII}}
\newcommand{\Mcnm}{M_{\rm CNM}}
\newcommand{\Mhalo}{M_{\rm halo}}
\newcommand{\nth}{n_{\rm th}}

\newcommand{\lc}{\ell_c}
\newcommand{\psfr}{\dot{\psi_{\star}}}
\newcommand{\Pmin}{P_{\rm min}}
\newcommand{\Pmax}{P_{\rm max}}
\newcommand{\Pgeo}{P_{\rm geo}}
\newcommand{\ncl}{n_{\rm cl}}
\newcommand{\CII}{[C\,{\sc ii}]\ }
\newcommand{\NHI}{{N_{\rm HI}}}
\newcommand{\HI}{H\,{\sc i}\,}
\newcommand{\cm}{\rm cm}
\newcommand{\Zsun}{Z_{\odot}}
\newcommand{\rhoc}{\rho_C}
\newcommand{\rhow}{\rho_W}

\newcommand{\Om}{\Omega_{\rm M}}
\newcommand{\Ol}{\Omega_{\Lambda}}
\newcommand{\Ob}{\Omega_b}
\def\avg#1{\left\langle#1\right\rangle}

\newcommand{\Lam}{\Lambda}

\newcommand{\hinv}{{h^{-1}}}
\newcommand{\mpc}{{\rm\,Mpc}}
\newcommand{\pc}{{\rm\,pc}}
\newcommand{\kpc}{{\rm\,kpc}}
\newcommand{\himpc}{\hinv{\rm\,Mpc}}
\newcommand{\hikpc}{\hinv{\rm\,kpc}}
\newcommand{\kms}{{\rm\,km\ s^{-1}}}
\newcommand{\kmsmpc}{{\rm\ km\ s^{-1}\ Mpc^{-1}}}

\newcommand{\yr}{{\rm yr}}
\newcommand{\Msun}{M_{\odot}}
\newcommand{\himsun}{\hinv{\Msun}}
\newcommand{\ltsim}{\lesssim}
\newcommand{\gtsim}{\gtrsim}

\newcommand{\beq}{\begin{equation}}
\newcommand{\eeq}{\end{equation}}

\slugcomment{DRAFT VERSION}

\shorttitle{\CII emission from high-z DLA galaxies}
\shortauthors{Nagamine et al.}

\begin{document}

\title{Detectability of \CII 158 $\mu$m emission from high-redshift galaxies: predictions for ALMA and SPICA}

\author{Kentaro Nagamine\altaffilmark{1}, Arthur M. Wolfe\altaffilmark{1}, 
Lars Hernquist\altaffilmark{2}}

\altaffiltext{1}{
Center for Astrophysics and Space Sciences, 
University of California, San Diego, MC 0424\\ 
9500 Gilman Dr., La Jolla, 92093-0424 \quad 
Email: knagamine@ucsd.edu, awolfe@ucsd.edu}

\altaffiltext{2}{Harvard University, 60 Garden Street, Cambridge, 
MA 02138}


\begin{abstract}

We discuss the detectability of high-redshift galaxies via \CII
158$\mu$m line emission by coupling an analytic model with
cosmological Smoothed Particle Hydrodynamics (SPH) simulations that
are based on the concordance $\Lambda$ cold dark matter (CDM) model.
Our analytic model describes a multiphase interstellar medium (ISM)
irradiated by the far ultra-violet (FUV) radiation from local
star-forming regions, and it calculates thermal and ionization
equilibrium between cooling and heating.  The model allows us to
predict the mass fraction of a cold neutral medium (CNM) 
embedded in a warm neutral medium (WNM).  
Our cosmological SPH simulations include a treatment of 
radiative cooling/heating, star formation, and feedback effects 
from supernovae and galactic winds.  
Using our method, we make predictions for the \CII luminosity 
from high-redshift galaxies which can be directly compared with 
upcoming observations by 
the {\it Atacama Large Millimeter Array} (ALMA) and the 
{\it Space Infrared Telescope for Cosmology and Astrophysics} (SPICA).  
We find that the number density of high-redshift galaxies 
detectable by ALMA and SPICA via \CII emission 
depends significantly on the amount of neutral gas 
which is highly uncertain.  
Our calculations suggest that, in a CDM universe, 
most \CII sources at $z=3$ are faint objects 
with $\Snu < 0.01$\,mJy.  
Lyman-break galaxies (LBGs) brighter than $\Rab=23.5$ mag 
are expected to have flux densities $\Snu = 1-3$\,mJy 
depending on the strength of galactic wind feedback.  
The recommended observing strategy for ALMA and SPICA 
is to aim at very bright LBGs or star-forming DRG/BzK galaxies.
\end{abstract}



\section{Introduction}

Up to now, the high-redshift galaxies 
detected in large numbers are observed by the radiation from 
stars in rest-frame UV to optical wavelengths.  
Such studies have been fruitful in delineating 
the properties of star-forming galaxies at $z\gtsim 3$ 
that are bright in UV; i.e., the Lyman-break galaxies (LBGs) 
\citep[e.g.,][]{Steidel99, Ade00, Shapley01, Iwata03, Steidel03,
Ouchi04a, Ouchi04b}.  
Within the hierarchical galaxy formation paradigm based on 
the cold dark matter (CDM) model \citep{Blumenthal84, Davis85}, 
it has been shown that the spectro-photometric properties of 
LBGs can be accounted for if we associate them with 
relatively massive galaxies ($M_\star \gtsim 10^{10} \Msun$) 
situated in large dark matter halos 
($M_{\rm halo}\gtsim 10^{12} \Msun$) 
\citep[e.g.][]{Mo96, Dave99, Nag02, Weinberg02, NSHM, 
Nachos2, Nachos3, Night06, Finlator06}.

However, observing the UV light emitted by the young stars 
does not directly tell us about how much gas there is 
in the galaxy.  
In order to obtain a full picture of galaxy formation, 
we need to find answers to the questions such as: 
1) How much neutral gas is available for star formation 
in high-redshift galaxies?
2) What is the baryonic mass 
(in particular, the neutral hydrogen mass) 
fraction as a function of halo mass, 
and how does it evolve as a function of redshift?
3) How does the volume-averaged neutral gas mass density 
evolve as a function of redshift? 
4) What are the main cooling and heating processes 
of the ISM in high-redshift galaxies?

One of the ways to detect the neutral hydrogen 
in high-redshift galaxies
is to search for damped Lyman-$\alpha$ absorbers 
\citep[DLAs,][for a review]{Wolfe05} in quasar absorption lines.  
DLAs are defined as quasar absorption systems with a neutral
hydrogen column density $\NHI>2\times 10^{20} \cm^{-2}$,
a threshold column density that effectively guarantees gas neutrality
at high redshifts \citep{Wolfe86, Wolfe05}.
Since DLAs are known to dominate the neutral hydrogen mass density 
at $z\gtsim 3$ \citep[e.g.,][]{Lan95, Storr00}, 
it is expected that they represent a significant reservoir of 
neutral gas for star formation.  
A large number of DLAs have been discovered at $z\gtsim 3$, 
and they have proven to be one of the best probes of neutral gas 
in high-redshift galaxies, complementary to 
the optical to infra-red (IR) observations.  
A recent search for DLAs in the Sloan Digital Sky Survey (SDSS) 
data archive yielded a large sample of over 500 DLAs 
at $z>2.2$ \citep*{Pro04, Pro05}.  
These observational results are beginning to put 
stringent constraints on theoretical/numerical models of 
galaxy formation.  
Furthermore, \citet{Wolfe03a} opened a new window for probing 
star formation in high-redshift DLA galaxies utilizing the 
C{\sc ii}$^*$ absorption lines, 
and this paper is motivated by their work.

The C{\sc ii}$^*$ absorption lines originate from the excited 
$^2P_{3/2}$ state in the ground 2s$^2$\,2p term of C$^+$. 
This is the same state that gives rise to \CII emission at 
$\lambda=157.74\,\mu$m through the $^2P_{3/2} \rightarrow ^2P_{1/2}$ 
fine structure transition. 
Because this is the dominant coolant of diffuse ISM 
at temperatures $T \le 5000$ K \citep{Dalgarno72, Tielens85, 
Wolfire95, Lehner04}, detection of \CII emission is
potentially another method for finding cold neutral gas 
at high redshifts \citep{Petrosian69, Loeb93}. 

\CII emission was first detected towards the inner regions of 
gas-rich spiral galaxies and starburst galaxies; i.e., dense 
star-forming gas irradiated by UV radiation from young star-forming 
regions near galactic centers 
\citep{Russell80, Crawford85, Stacey91, Carral94}. 
In this case, the brightness of the emission line suggests 
that it is produced in warm ($T > 200$\,K), 
dense ($n_{\rm H} > 10^3$\,cm$^{-3}$) photodissociated regions (PDRs) 
with pressures $P/k_B=10^4 - 10^7$\, K\,cm$^{-3}$ 
at the interface between giant molecular clouds and 
fully ionized media. 
The \CII luminosity accounts for $0.1 - 1$\% of the total 
far-IR luminosity of the nuclear regions.

However, later space-based observations with 
the Long Wavelength Spectrometer (LWS) on-board ESA's 
Infrared Space Observatory (ISO) have shown that, 
for quiescent late-type galaxies like NGC6946, 
the \CII emission from extended diffuse gas over the entire 
disk could also be significant compared to that from dense 
compact star-forming gas \citep[e.g.,][]{Madden93, Malhotra97, 
Leech99, Malhotra01, Contursi02}. 
The diffuse cold components of the ISM seems to be at least 
as important sources of the \CII emission as are compact regions 
\citep[e.g.,][]{Sauty98, Pierini99, Pierini01, Contursi02}. 

More recently, \CII emission has also been detected in the 
highest redshift SDSS quasar at $z=6.42$ \citep{Maiolino05}.
Quasars are considered to host significant amounts of 
molecular gas ($M_{\rm gas}\approx 10^{11}\Msun$) and 
dust ($M_{\rm dust}\approx 10^8\Msun$), and are therefore 
forming stars at a significant rate of $SFR \gtrsim 1000\,\Msun\,
\yr^{-1}$ \citep[e.g.,][]{Omont96, Omont01, Omont03, Carilli01,
Walter03, Beelen04, Carilli04}. 
Given this large amount of gas contained in them, 
they should be very bright in \CII luminosity, 
allowing them to be easily detected even at very high redshift.  
However, bright quasars have much lower comoving space density
than LBGs and are not representative of the entire galaxy population; 
they are special sources hosting supermassive black holes, 
and may be in unusual dynamical states 
if their activity is triggered by mergers \citep[e.g.,][]{Hopk05}. 

The detection of \CII emission from normal 
(i.e., not quasars or active galactic nuclei [AGN]) 
high-redshift star-forming galaxies would be a more 
{\it direct} way than the absorption line technique to
find out about the amount of neutral gas 
available for star formation. 
Such detections of \CII emission from high-redshift galaxies, 
for example, by ALMA and SPICA,
would set an important constraint on the theory of galaxy formation.  
In particular interferometric maps of \CII emission 
would reveal the spatial dimensions of the DLAs, 
a property that so far has eluded detection. 
By combining these measurements with the velocity widths 
of the \CII emission lines, it will be possible to 
infer the dark matter masses of these objects, 
and the luminosity of the \CII emission lines 
would tell us the heating rate of the gas.

In this paper we study the detectability 
of high-redshift DLA galaxies 
by coupling an analytic model 
with cosmological SPH simulations 
that are based on the concordance $\Lambda$CDM model, 
and make predictions for upcoming observations 
by ALMA and SPICA.
We also compare the computed \CII luminosity function 
with the observed LBG luminosity function, 
and discuss the connection between DLA galaxies and LBGs. 
\citet{Stark97} discussed the potential measurement of 
\CII emission from high-redshift galaxies, 
but he made various assumptions for the \CII luminosity function 
and the evolution of the characteristic luminosity 
based on the local observations and numerical simulations, 
and focused more on the observing time and conditions
rather than the intrinsic nature of high-redshift galaxies.  
The present study extends his original work, 
and is intended to be more physically motivated 
by using full cosmological hydrodynamic simulations 
which simulate structure formation from early epochs 
to the present time. 
We note that \citet{Suginohara99} studied the detectability 
of atomic emission lines of carbon, nitrogen, and oxygen, 
but they focused on the intensity fluctuations 
owing to sources at $z\sim 10$ rather than 
the detection of individual objects.


\section{Cosmological Hydrodynamic Simulations}
\label{sec:simulation}

The cosmological Smoothed Particle Hydrodynamics (SPH) simulations
that we use in this paper were carried out using {\small GADGET-2}
\citep{Springel05}. This code adopts the novel `entropy-formulation'
of SPH \citep{SH02}, and includes `standard' physical processes 
such as radiative cooling and heating by 
a uniform UV background of modified \citet{Haardt96} form 
\citep[see][]{Dave99}, star formation, supernova feedback, 
as well as a phenomenological model for feedback by galactic winds. 
The latter allows us to examine the effect of galactic outflows.  
(For now, we do not account for feedback associated with 
black hole growth, as in, e.g., \citet{Springel05b, Springel05c, 
DiMatteo05}, but plan to investigate consequences of 
this mechanism in due course.)  
The ionization equilibria of hydrogen and helium are solved 
assuming an optically thin limit throughout the simulation box. 
We will discuss this point in Section~\ref{sec:CNMgas}.  
We utilize a series of simulations of varying box-size and 
particle number to isolate the impact of numerical resolution 
on our results.  
The important parameters of the simulation runs are 
summarized in Table~\ref{table:simulation}.  
The same simulations were used for the study of 
the cosmic star formation history \citep{SH03b, Nachos1}, 
LBGs at $z=3-6$ \citep{NSHM, Night06, Finlator06}, 
damped Lyman-$\alpha$ systems \citep{NSH04a, NSH04b, Nag05b}, 
massive galaxies at $z=2$ \citep{Nachos2, Nachos3}, and 
the intergalactic medium \citep{Furla03, Furla04d, Furla05}.  
The adopted cosmological parameters of all runs are 
$(\Om,\Ol,\Ob,\sigma_8, h)= (0.3, 0.7, 0.04, 0.9, 0.7)$, 
where $h \equiv H_0 /(100 \kmsmpc)$ is the Hubble constant.  
We also use the notation $h_{70} \equiv h / 0.7$.

The initial conditions for our simulations are set by tiny density
fluctuations as motivated by inflationary theories, and the
calculations follow the development of structure using the laws of
gravity and hydrodynamics.  Therefore, in principle the cosmological
simulations allow us to model galaxy evolution in a physical manner.
However, even with current state-of-the-art computers, we still lack
the computational resources to directly simulate the details of the
ISM dynamics below $\sim 1$\,kpc while at the same time solving for
large-scale structure formation on tens of megaparsecs.  
Consequently, the simulations that we utilize in this paper 
do not directly resolve the multiphase nature of the ISM 
with temperature $T<10^4$\,K, but keep track of the dynamics 
of the hot ionized medium (HIM) with $T>10^4$\,K 
that is pressurized by the supernova feedback in star-forming regions, 
as we will describe later in Section~\ref{sec:SFgas}. 
The mass of the cold gas with $T<10^4$\,K is then estimated 
with a sub-resolution ISM model as described in detail
by \citet{SH03a, SH03b}. In this sub-resolution ISM model, 
the hot and cold phase is assumed to be in pressure equilibrium 
\citep{McKee77}. The mass in high-density star-forming gas is 
dominated by the cold gas with $T<10^4$\,K, and the volume is 
dominated by the hot component. 

In the following sections, we will further supplement the simulation 
with a two-phase sub-resolution model in order to divide 
the cold gas with $T<10^4$\,K in the simulation 
into a cold neutral medium (CNM, with $T \sim 80$\,K and 
$n\sim 10\, \cm^{-3}$) and a warm neutral medium (WNM, with 
$T \sim 8000$\,K and $n\sim 0.1\, \cm^{-3}$). 
Therefore, in principle we are picturing a 3-phase medium 
with HIM, WNM, and CNM \citep[cf.][]{McKee77}. 

Within this multiphase ISM picture, 
\citet{Wolfire03} estimated that the scale $\lp$ 
on which turbulent pressure begins to dominate the 
thermal pressure is $\sim 215$\,pc for the WNM, 
based on the comparison of turbulent velocity dispersion 
and thermal velocity dispersion. 
Also the model of the vertical distribution of the \HI 
by \citet{Dickey00} suggests that the FWHM of 
the vertical gas distribution of the WNM is $\sim 500$\,pc 
\citep{Wolfire03}. 
These two estimates indicate that the volume occupied 
by the warm ionized medium and the WNM could be substantial
compared to that of the HIM in an ISM with a scale $\sim 1$\,kpc. 
Furthermore recent high-resolution numerical simulations 
of a turbulent ISM suggest that the volume fraction of WNM 
could be comparable to that of HIM 
\citep{Kritsuk02a, Kritsuk02b, Kritsuk04}. 
Since we are focusing on high-density ISM
where the mass and density is dominated by the neutral component, 
here we restrict ourselves to the two-phase picture of 
the CNM and WNM, and in the following neglect the volume 
occupied by the HIM for the simplicity of the present 
sub-resolution model. 
Obviously this is an over-simplification of the actual 3-phase ISM, 
and could lead to an over-estimate of the volume fraction of CNM
and thereby to an overestimate of the total \CII emission 
of galaxies.
Therefore, the predicted amount of CNM gas based on our 
simplified model is more likely to be an overestimate rather 
than an underestimate.
By the construction of our model, the total amount of CNM 
determined in our simulation cannot exceed the total amount of 
cold neutral gas (with $T<10^4$\,K, i.e., CNM+WNM) 
given by the simulation.


\section{The Model}
\label{sec:model}

In this section, we describe a simple analytic model to estimate 
the mass fraction of the CNM for given mean gas density, 
FUV radiation intensity, metallicity, and dust-to-gas ratio. 
The notation that we use hereafter in the paper is summarized 
in Table~\ref{table:notation}. As noted earlier, our work is largely
motivated by that of \citet{Wolfe03a}. 
These authors recently opened a new window for probing 
star formation in high-redshift galaxies utilizing the 
C{\sc ii}$^*$ absorption line observed in DLAs. 
The idea is to use this absorption line to infer the cooling rate 
owing to the \CII fine structure line, which in turn tells us 
about the heating rate caused by the FUV radiation field 
from nearby star formation.
One can then estimate the projected star formation rate ($\psfr$) 
in DLAs as the heating rate is expected to be proportional 
to the local star formation rate (SFR). 

\citet{Wolfe03a} considered a two-phase model in which 
the CNM and WNM are in pressure equilibrium. 
Subsequently \citet{Wolfe03b} ruled out a solution in which 
C{\sc ii}$^{*}$ absorption arises in the WNM, 
as it overpredicts the observed bolometric extragalactic 
background radiation. 
\citet{Howk05} showed directly that C{\sc ii}$^{*}$ absorption 
in one DLA cannot arise in the WNM.  
Since C{\sc ii}$^{*}$ absorption arises from the same 
$^2P_{3/2}$ state which gives rise to the \CII emission, 
we assume that the CNM is the dominant source for the \CII emission 
in this paper.


\subsection{Set-up \& Key Equations}

Consider a ISM with mean density $\rho_0$ and volume $V_0$ 
in the vicinity of a star-forming region. 
This interstellar cloud consists of CNM and WNM 
that are in thermal equilibrium within each cloud, 
and the two have density $\rho_C$ and $\rho_W$, 
volume $V_c$ and $V_W$, respectively.  
We also denote the volume fraction occupied by CNM 
by $f_V = V_C / V_0$, and the mass fraction of the CNM by $f_M$. 
Given these definitions, the following simple relations 
immediately follow:
\begin{eqnarray}
\rho_C V_C + \rho_W V_W &=& \rho_0 V_0, \quad\quad ({\rm mass\: conservation}) \label{eq:massconsv}\\
V_C + V_W &=& V_0, \quad\quad\quad ({\rm volume\: conservation}) \label{eq:volconsv}\\
f_M &=& \frac{\rho_C V_C}{\rho_0 V_0} = \frac{\rho_C}{\rho_0} f_V. \label{eq:fm}
\end{eqnarray}
As mentioned above, in Equations~(\ref{eq:massconsv}) \& 
(\ref{eq:volconsv}), the volume of the HIM is ignored, 
and we are assuming that the volume fraction of the WNM is 
fairly substantial. 
Eliminating $V_W$ from Equations~(\ref{eq:massconsv}) \& 
(\ref{eq:volconsv}) gives
\begin{eqnarray}
f_V = \frac{V_C}{V_0} &=& \frac{\rho_0 - \rhow}{\rhoc - \rhow}
=\frac{1-(\rhow / \rho_0)}{\left(\frac{\rhoc}{\rhow}-1\right) {\rhow \over \rho_0}}, \quad {\rm and} \label{eq:fv}\\
\rho_0 &=& (1 - f_V) \rhow + f_V \rhoc. 
\label{eq:rho0}
\end{eqnarray}
Equation~(\ref{eq:rho0}) can also be obtained from 
Equation~(\ref{eq:massconsv}) by dividing both sides by $V_0$.

Combining equations (\ref{eq:fm}), (\ref{eq:fv}), \& (\ref{eq:rho0}) gives
\beq
f_M = \frac{1}{\left ( \frac{1-f_V}{f_V}\right ) \left (\frac{\rhow}{\rhoc}\right ) + 1} = \frac{1-(\rhow / \rho_0)}{1-(\rhow / \rhoc)}. 
\label{eq:fm2}
\eeq
Therefore, if we have estimates for $\rhoc$/$\rhow$, and $\rhow$/$\rho_0$
(or equivalently $\rho_C$/$\rho_W$, and $f_V$), 
then we can compute 
the mass fraction of the CNM using Equation~(\ref{eq:fm2}).
Note that knowing the volume fraction of the CNM, $f_V$, is equivalent 
to constraining the mean density $\rho_0$ if $\rhoc$ and $\rhow$ are
known, and the two quantities ($f_V$ and $\rho_0$) are related to 
each other  via Equation~(\ref{eq:fv}) or (\ref{eq:rho0}).


\subsection{Phase Diagrams}
\label{sec:phase}

In order to identify the CNM and WNM densities, 
we compute the two-phase structure of a multiphase ISM 
by solving the equations of thermal and ionization equilibrium 
for a given gas density, FUV radiation field intensity, metallicity, 
and dust-to-gas ratio with the numerical techniques and 
iterative procedures outlined in \citet{Wolfire95} and \citet{Wolfe03a}.  
We describe the further details of our method in Appendix A.

Solving for the thermal balance results in a characteristic shape 
of pressure curves as a function of gas density 
as shown in the top panels of Figure~\ref{fig:phase}
with a local minimum ($P=\Pmin$) and maximum ($P=\Pmax$) 
defined by $\partial(\log P)/\partial(\log n) = 0$.
In this figure, the pressure curves are shown as a 
function of gas density for different input projected SFR 
per unit physical area ($\psfr$) and metallicity ($\log (Z/\Zsun)$).
Our full grid of models covers the range 
$\log \psfr\, [\Msun\, \yr^{-1} \kpc^{-2}] = 
-4.0, -3.5, -3.0, -2.5, -2.0, -1.5, -1.0, -0.5,$ \& 0.0, and 
$\log(Z/Z_\odot) = -2.5, -2.0, -1.5, -1.0, -0.5,$ \& 0.0.
Two things can be read off from this figure: 
(1) When $\psfr=0$ and only the UV background radiation is present, 
the resulting pressure curve appears almost identical to the case
of $\log \psfr = -4.0$; i.e., the impact of the UV background is 
negligible when $\log \psfr \gg -4.0$.
(2) As $\psfr$ (or equivalently, the incident FUV radiation intensity) 
increases, the pressure maxima and minima move towards 
the upper right corner. 
For densities between those corresponding to $\Pmax$ and $\Pmin$,
the cooling curve is relatively insensitive to changes 
in temperature $T$.
Therefore, at a given density, $T$ undergoes a large increase 
when the cooling rate rises to match the increase in heating rate, 
thereby increasing the pressure.  
Conversely, outside this range in densities, 
the cooling curve increases rapidly with increasing $T$ 
and as a result the pressure is insensitive to increases
in heating rate.
Also, as the metallicity decreases, metal-line cooling becomes 
inefficient, and the density and temperature have to be raised 
for a given heating rate, and therefore the equilibrium pressures go up. 

It is known that the two thermally stable phases can coexist 
over a narrow range of pressures $\Pmin < P < \Pmax$, where 
$\partial(\log P)/\partial(\log n) > 0$.  Here, we follow the work 
by \citet{Wolfire95} and \citet{Wolfe03a}, and assume that the 
equilibrium gas pressure equals the geometric mean pressure 
$P = \Pgeo \equiv \sqrt{\Pmin \Pmax}$ for identifying $\rhoc$ and $\rhow$, 
as indicated by the open triangles in the phase diagrams shown 
in the top panels of Figure~\ref{fig:phase}.  
In passing, we note that \citet{Kritsuk02a, Kritsuk02b} have shown, 
using high resolution numerical simulations,
that the thermally unstable turbulent ISM settles down 
as a multiphase medium, where the majority of the masses 
(for both CNM and WNM) lie close to the local minimum pressure $\Pmin$.

For each equilibrium curve, the model of \citet{Wolfe03a} gives 
the \CII luminosity per hydrogen atom, $\lc$, as a function of 
gas density as we describe in Appendix A. 
The value of $\lc$ for the CNM can be determined at the CNM density 
as shown by the open triangles in the bottom panels of 
Figure~\ref{fig:phase}. 
The value of $\lc$ increases with increasing metallicity 
as the heating of the gas (and correspondingly the cooling rate 
to match that) becomes more efficient via 
the grain photoelectric heating effect. 
At low metallicity (e.g., $\log(Z/\Zsun)=-2.5$, 
the bottom left panel of Figure~\ref{fig:phase}), 
the value of $\lc$ begins to decrease at high densities, 
because grain photoelectric heating becomes more inefficient 
owing to low dust content, and 
the heating becomes dominated by X-rays. 
The X-ray heating rate decreases with density since 
it becomes less efficient with increasing neutral gas fraction, 
which increases with density.   
In the case of $\log (Z/\Zsun)= -0.5$, 
the value of $\lc$ continues to increase with increasing density 
as the grain photoelectric heating remains efficient. 
The flattening of the curves at low density occurs when 
the cosmic microwave background (CMB) photons dominate 
the population rates of the fine-structure states.
Note, the flattening is absent from the figure at very low metallicity
since the population of the $^2P_{3/2}$ state is proportional to 
metallicity when the CMB dominates.


\subsection{Mass fraction of CNM: $f_M$}
\label{sec:CNMfrac}

Figures~\ref{fig:fm} and \ref{fig:pmin} illustrate the properties 
of our multiphase ISM model.  Figure~\ref{fig:fm} shows 
$\rhoc$, $\rhow$, the ratio of the two, and $f_M$ as functions of 
projected SFR $\psfr$ for different metallicities, while 
Figure~\ref{fig:pmin} shows $\Pmin$ as a function of $\psfr$ for 
different metallicities. 
Figures~\ref{fig:fm}a, \ref{fig:fm}b, and \ref{fig:pmin} demonstrate 
that $\rhow$, $\rhoc$, and $\Pmin$ are smooth 
monotonically increasing functions of $\psfr$ 
as expected from Figure~\ref{fig:phase}; 
as $\psfr$ increases, the equilibrium pressure $\Pgeo$ increases 
as does the equilibrium density. 
For a given $\psfr$, as the metallicity decreases, 
the equilibrium density increases for cooling to match the heating rate.  
The CNM and WNM density scale roughly similarly, 
therefore the ratio $\rhow / \rhoc$ is relatively flat 
as a function of $\psfr$ as shown in Figure~\ref{fig:fm}c 
except for $\log(Z/\Zsun)<-2.0$ and $\log \psfr > -1.0$.  
As a result, $f_M$ is a monotonically {\em decreasing} function 
of $\psfr$. 
Here, the values of $f_M$ were computed by Equation~(\ref{eq:fm2}) 
assuming a constant value of $f_V = 7\times 10^{-4}$, which is 
motivated by the following argument. 

Consider a column of length $L$ and a cross section of 
unit physical area through the ISM. 
Let $f_A$ denote the area covering fraction of the CNM, 
and $\ncl$ the number density of the CNM cloud, 
and $R$ the physical radius of each CNM (spherical) cloud. 
Then, 
\begin{eqnarray}
f_V &=& \frac{4\pi}{3}R^3\, \ncl, \quad {\rm and}\\
f_A &=& \pi R^2\, \ncl\, L.
\end{eqnarray}
Combining the above two equations and assuming typical values of 
$f_A \sim 0.5$, $L \sim 1 \kpc$, and $R\sim 1 \pc$ gives 
\begin{eqnarray}
f_V &=& 1.33\, f_A\, \frac{R}{L} 
\sim 1.33 \times 0.5 \times \frac{1\pc}{1\kpc} = 7\times 10^{-4}, 
\label{eq:fv2}
\quad {\rm and}\\
\ncl &=& \frac{f_A}{\pi R^2\, L} = 1.6\times 10^{-4}~~ {\rm pc}^{-3}. 
\end{eqnarray}
We take $f_A \sim 0.5$ because \citet{Wolfe03a} find
CNM gas in roughly half of the observed DLAs. 
If we instead adopt $L=100$ pc, then 
$f_V = 7\times 10^{-3}$ and $\ncl = 1.6\times 10^{-3}$ pc$^{-3}$, 
which are in better agreement with the values 
adopted by \citet{McKee77}. 
However the latter values result in $f_A$ greater than unity. 

It is not surprising that the value of $f_V$ is very small, 
as the volume occupied by CNM clouds should be much smaller 
than that of the WNM. 
If we take Equation~(\ref{eq:fv2}) and compute the values of $f_M$ 
using Equation~(\ref{eq:fm2}), we obtain Figure~\ref{fig:fm}d. 
The exact values of $f_V$ and $f_M$ are not important here, 
because when we later couple this model with cosmological 
simulations, we do {\it not} assume a constant value of $f_V$. 
Instead, we take the mean density of the ambient ISM $\rho_0$
directly from our hydrodynamic simulations and  
compute the value of $f_M$ for each gas element 
in the simulation using Equation~(\ref{eq:fm2}). 

In summary, the procedure to obtain the CNM mass fraction in our model 
is as follows:
\begin{enumerate}
\item Compute a grid of equilibrium curves in the phase space 
of pressure and density for a range of projected star formation rates 
and metallicities (top panels of Figure~\ref{fig:phase}).
\item For each equilibrium curve, identify $\rhoc$ and $\rhow$ at 
$\Pgeo = \sqrt{\Pmin \Pmax}$.
\item Use Equation~(\ref{eq:fm2}) and the inferred values of $\rhoc$ 
and $\rhow$ to obtain $f_M$.  
Here we take the mean gas density $\rho_0$ directly from 
the result of cosmological simulations, and compute
the amount of CNM gas for each dark matter halo as we show in 
Section~\ref{sec:CNMgas}. 
The model of \citet{Wolfe03a} gives the \CII luminosity per H atom 
$\lc$ as a function of $\psfr$ (see Appendix A), and we describe
how we compute the \CII luminosity of each halo in 
Section~\ref{sec:CIIlum}. 
\end{enumerate}


\section{Results from Cosmological Simulations}

\subsection{Star-forming gas in simulations}
\label{sec:SFgas}

Let us first look at the range of gas pressure and density 
that our cosmological hydro simulations cover.  
Figure~\ref{fig:pn} shows the distribution of cosmic gas 
in our cosmological simulations in the plane of density vs. pressure. 
In the simulation, the gas that has density higher than the 
threshold density $\nth = 0.13$\,cm$^{-3}$ is treated 
as a multiphase medium and is able to form stars, 
as described by the sub-resolution ISM model of \citet{SH03a, SH03b}. 
We consider this star-forming gas as the ambient ISM 
(with mean density $\rho_0$) that hosts CNM and WNM.
The magenta {\it dashed} curve to the right of 
the SF threshold density is the analytic fit to the effective 
equation of state adopted in the simulation \citep{Robertson04}.
{\em We stress that in Figure~\ref{fig:pn} we are plotting 
$\rho_0$ vs. $P$ for the simulations.} 
The star-forming gas in the simulation is pressurized 
by heating owing to supernova feedback, and has a higher pressure 
than isothermal ($P\propto n$) gas as traced by the magenta
{\it dashed} line. 
The Q5, D5, and G5 runs also include a treatment of feedback 
owing to galactic winds.  
In this approach, a fraction of supernova feedback energy 
is given to star-forming gas particles as kinetic energy 
and momentum in random directions, and some gas particles are 
ejected from star-forming regions owing to winds. 
Therefore, one can see that the spread of the gas distribution 
around the effective equation-of-state is larger 
in the Q5, D5, and G5 runs than in the O3 (no wind) run. 
The distribution of gas in the Q5, D5, and G5 runs is 
roughly the same. 

The three sets of two lines indicate the $\Pmin$ obtained from 
the pressure curves for metallicities of, 
from upper right to bottom left, 
$\log (Z/\Zsun) = -2.5$ (black), $-1.0$ (blue), and $0.0$ (red), 
as a function of 
$\rhoc$ (the CNM density, shown in solid lines) and 
$\rhow$ (the WNM density, shown in {\it dot-dashed} lines) 
at $\Pgeo$, respectively. 
We stress again that in Figure~\ref{fig:pn} we are plotting 
$\rho_0$ vs. $P$ for the simulations, whereas the analytic model 
calculation results show $P$ vs. $\rhoc$ and $\rhow$. 
The criteria that we impose for the simulated gas
to qualify for hosting CNM are $P>\Pmin$ and $\rho_0 > \rhow$ 
(see Equations~[\ref{eq:fv}] and [\ref{eq:fm2}]). 
The condition $\rhoc > \rhow$ is guaranteed by the construction 
of the model. 

The highest gas densities $\rho_0$ attained in our simulations are 
$n\sim 300\,\cm^{-3}$ in the Q5 run, and $n\sim 100\,\cm^{-3}$ in 
other runs. 
Most of the star-forming gas in the simulations has a higher 
pressure than $\Pmin$ in all runs, but we need to divide 
the gas into different metallicity ranges to perform 
a more detailed comparison.

In Figure~\ref{fig:pn_metal}, the gas in the Q5 run is divided into 
different metallicity ranges, and is compared to the corresponding 
values of $\Pmin$ for the same metallicity range. 
It is seen that most of the star-forming gas, i.e., 
gas with $\nth \ge 0.13$\,cm$^{-3}$,  in the simulation is 
relatively metal-rich, and there is almost no star-forming gas with 
$\log (Z/\Zsun) < -2.5$ that satisfies $P>\Pmin$. 
Since the simulation does not resolve 
the high density CNM directly, it is not a problem that
the diluted density of the simulated gas does not fall 
in-between the range $\Pmin<P<\Pmax$ for the CNM densities $\rhoc$.


\subsection{CNM gas in dark matter halos}
\label{sec:CNMgas}

Before we compute the CNM mass in each dark matter halo, 
there are several steps in processing the simulation data. 
First, we identify dark matter halos by applying a conventional 
friends-of-friends algorithm to the dark matter particles
in each simulation. After dark matter halos are identified, 
we set up a 3-dimensional cubic grid centered at the center 
of each dark matter halo covering the entire halo, 
with grid cell-size equal to the gravitational softening length 
of each simulation. 
Then, the gas mass, H\,{\sc i} mass, metal mass, and 
star formation rate of each gas particle is smoothed 
over a spherical region of grid-cells, weighted by the SPH kernel. 
We now have all of the above quantities for each grid cell. 
These procedures are the same as adopted by \citet{NSH04a, NSH04b}. 

We now compute the CNM mass contained in each dark matter halo. 
Given the four quantities $(\rho_0, P, \psfr, Z/\Zsun)$ 
in each grid cell that covers the dark matter halo
and the two quantities $(\rhoc, \rhow)$ obtained from 
the phase diagram shown in Figure~\ref{fig:phase}, 
we can calculate the value of $f_M$ for each grid cell 
using Equations~(\ref{eq:fm2}). 
Here, only those cells that satisfy $P>\Pmin$ and $\rho_0 > \rhow$ 
(see Equations~[\ref{eq:fv}] and [\ref{eq:fm2}]) 
are considered to host CNM. 
Figure~\ref{fig:fm_cont} shows the volume-averaged value of $f_M$ 
for each halo, $\avg{f_M}$, where all the grid cells in each halo
are equally averaged over and each point in this plot represents
one halo. Contours are used to represent the scatter and avoid 
a saturation of points in the figure.   
The values of $\avg{f_M}$ are in the range $0.5 - 0.8$ 
for relatively massive halos with $\Mhalo > 10^{11}\himsun$. 
The O3 run has higher values of $\avg{f_M}$ owing to 
the absence of winds. The rapid decline of $\avg{f_M}$ values 
at $10 < \log \Mhalo < 12$ for the D5 and G5 runs 
owes to lower resolution in simulations with large box sizes.  
We consider that a true physical decline of $\avg{f_M}$ should occur
at $\log \Mhalo \sim 8.5$, because at this mass-scale 
the DLA cross section decreases rapidly in the highest resolution 
runs as shown in Figures~2 \& 3 of \citet{NSH04b}, 
and hence the CNM mass fraction is expected to decrease as well. 

Using the value of $f_M$ computed for each grid cell that 
covers the dark matter halo, 
we calculate the total CNM mass of each halo according to
\beq
\Mcnm = \Sigma_i f_M \rho_0 V_0, 
\label{eq:Mcnm}
\eeq
where $\rho_0$ and $V_0$ are the gas density and the volume of 
each grid cell, and the index $i$ runs through all grid cells for 
each halo.  The resulting $M_{\rm CNM}$ for each halo is shown 
in Figure~\ref{fig:nhmass}. The shaded contours are for the CNM, and
the black contour without the shade is the total neutral hydrogen
mass in the halo.  As we described in Section~\ref{sec:simulation}, 
ionization equilibria of hydrogen and helium are solved assuming
an optically thin limit across the simulation box. However, in 
high density star-forming regions, the recombination time-scales are
expected to be shorter than the ionization time-scales, rendering
some reliability in our estimate of the total neutral hydrogen mass
in each halo (see Section~\ref{sec:conclusion} for the prospects of 
improving upon this in the future). 
The CNM mass in our model cannot exceed the black contour 
by construction. 
The {\it short-dashed, dash-dot, long-dashed} lines indicate the 
5\%, 1\%, and 0.5\% of the halo mass, respectively. 

In the case of the O3 (no wind) run, the bulk of gas 
(both total neutral mass and the CNM mass) has mass fractions 
in-between 1 to 5\% for most of the halos. 
These values are very close to that of the 
disk mass fraction 0.05 adopted by \citet{Mo98}. 
The most massive halos in the O3 run have slightly lower 
neutral mass fractions ($\sim 3$\%) than this value. 
When galactic wind feedback becomes strong (Q5, D5, and G5 runs), 
the total neutral mass fraction goes down to $\sim 1$\%, 
and the CNM mass fraction to even lower values ($0.1 - 1$\%). 
This is because the gas is heated and ejected owing to winds, 
resulting in a lower neutral fraction. 
Similarly to Figure~\ref{fig:fm_cont}, the rapid decline 
of $\Mcnm$ values at $10 < \log \Mhalo < 12$ for D5 and G5 run 
owe to lower resolution in simulations with large box-sizes, 
and we consider that a true physical decline of $\Mcnm$ 
should also occur at $\Mhalo \sim 10^{8.5}\,\himsun$.


\subsection{\CII emission from CNM}
\label{sec:CIIlum}

Having obtained the CNM mass of each grid cell computed 
in the previous section, 
we can now compute the \CII luminosity of each halo 
by performing a similar sum to Equation~(\ref{eq:Mcnm}) 
with an additional multiplication of \CII luminosity per H atom, 
$\lc$, as follows:
\beq
\Lcii = \Sigma_i \ \lc(Z, \psfr)\, \frac{f_M\, \rho_0\, V_0}{\mu\, m_{\rm H}}, 
\eeq
where $\mu$ is the mean molecular weight and $m_{\rm H}$ is the 
hydrogen mass. Notice that $\lc$ is a function of metallicity $Z$ and 
$\psfr$ of each grid cell. 
The resulting \CII luminosity of each halo is shown in 
Figure~\ref{fig:cii}. Each point in this diagram represents
a halo, but we use contours in order to avoid saturation in the 
scatter plot.  
The {\it long-dashed} and {\it short-dashed} lines 
correspond to the following scaling relationships:
\beq
\Lcii = C_1 \left( \frac{\Mhalo}{10^{12}\,\himsun} \right),
\label{eq:Lcii}
\eeq
where $C_1 = 10^{41}$ and $10^{40}$\,erg\,s$^{-1}$, respectively.
Most of the halos except for the least massive ones in the `no-wind' 
(O3) run follow $C_1 = 10^{41}$\,erg\,s$^{-1}$ well, 
and the `strong wind' (Q5, D5, G5) runs are better characterized 
by $C_1 = 10^{40}$\,erg\,s$^{-1}$. The decrease of $\Lcii$ in the least
massive halos in each run occur owing to same reasons as mentioned for 
Figure~\ref{fig:fm_cont}.  

Another way to characterize the distribution of \CII luminosity of halos
is to look at the cumulative luminosity function, which is shown 
in Figure~\ref{fig:cii_lf}. As expected from Figure~\ref{fig:cii},
the `no-wind' (O3) run is brighter than the `strong wind' 
(Q5, D5, and G5) runs by about an order of magnitude. 
The bright-end of the O3 and Q5 run is limited by cosmic variance 
owing to small simulation box-sizes, and the faint-end of the D5 and G5
runs are limited by the low resolution in large box-size simulations.  
The total luminosity function can be obtained by interpolating the 
results of the three runs when they overlap. We will perform
such interpolation in the next section and in Figure~\ref{fig:snu}. 


\subsection{Comparison with LBG Luminosity Function}

The connection between DLA galaxies and LBGs is of significant 
interest. The recent measurements of cross-correlation between DLAs
and LBGs \citep{Gawiser01, Ade03, Bouche03, Bouche04, Cooke06}
suggest that the typical DLA halo mass could be similar to that
of the LBGs' ($\Mhalo \simeq 10^{12}\,\Msun$), and that DLAs could 
be strongly correlated with LBGs. Since LBGs contribute a 
significant fraction of star formation rate density 
at $z\simeq 3$ and DLAs dominate the total H\,{\sc i} gas density
at the same redshift, it is natural to expect that the two systems
have some connection with each other. In this case, the energy 
input source for the heating discussed in Section~\ref{sec:model} 
would be the central LBGs in the halo rather than the {\em in situ}
star formation within DLAs \citep{Wolfe06}.  

In order to compare the computed \CII luminosity function with 
the observed LBG population, we show in Figure~\ref{fig:cii_lf} 
the observed number density of LBGs at $z=3$ 
with magnitudes $\Rab < 25.5$ by \citet{Ade00} and \citet{Ade03} 
with a yellow shaded region. 
In addition, the following simple scaling laws and 
the observed cumulative luminosity function of LBGs 
can be used to obtain the magenta {\it dot-long-dashed} curve. 
Using the results of population synthesis calculations, 
\citet{Nag05b} obtained a relationship between $\Rab$ magnitude 
and halo mass $\Mhalo$ as 
\beq
\Rab = -2.5 \log \Mhalo + C_3,
\label{eq:Rab_scale}
\eeq 
where $\Mhalo$ is in units of $\himsun$ and 
$C_3 = 55.03$ (O3 run) and 57.03 (Q5 run). 
Inserting this equation into Equation~(\ref{eq:Lcii}) gives 
\beq
\log \Lcii = -0.4\,(\Rab - C_3) + \log C_1 - 12.
\label{eq:Lcii_scale}
\eeq
For the case of the O3 run, this results in 
$\log \Lcii = -0.4\,\Rab + 51.01$. 
Next, we compute the cumulative luminosity function of LBGs
using a Schechter fit with parameters $(m^*, \alpha, 
\Phi^*[h^3\mpc^{-3}]) = (24.54, -1.57, 4.4\times 10^{-3})$ 
obtained by \citet{Ade00}. 
We then convert the abscissa of this cumulative function
using Equation~(\ref{eq:Lcii_scale}) into \CII luminosity. 
As a result, the two magenta {\it dot-long-dashed} curves 
in Figure~\ref{fig:cii_lf} are obtained: 
the brighter one on the right is for the O3 run scaling, 
and the fainter one on the left is for the Q5 run scaling.   
We also indicate the magnitudes $\Rab = 23.5$, 25.5, 30.0, and 36.0 
on this curve with filled triangles for the O3 run, 
which correspond to $\log \Lcii = 41.61$, 40.81, 39.01, 
and 36.61, respectively. 
The agreement between the two magenta curves and 
the actual simulation results is not perfect, but 
the two magenta curves lie in-between the `no-wind' (O3) result 
and the `strong wind' (Q5, D5, and G5) results, 
which we consider reasonable given the crudeness
of the above scaling relationships.


\subsection{\CII flux density}

The \CII flux density of each halo can be computed by 
\beq
\Snu = \frac{(1+z)\,\Lnu}{4\,\pi\, {d_L}^2} = 
\frac{(1+z)}{4\, \pi\, {d_L}^2}\, \frac{\Lcii}{\Dnu},
\label{eq:snu}
\eeq
where $\Dnu = \nucii (v_c / c)\, \eta$ is the line width, 
$\nucii = 1897$\,GHz is the rest-frame frequency of the \CII emission line, 
and $v_c$ is the circular velocity of the dark matter halo
at a radius of overdensity 200, 
\begin{eqnarray}
v_c &\equiv& \left( \frac{G\Mhalo}{R_{200}}\right)^{1/2}
= \left[ G \Mhalo^{2/3} \left(\frac{4\pi}{3}\bar{\rho}\,200\right)^{1/3} \right]^{1/2} \\ 
&=& 123.5\, \left( \frac{\Mhalo}{10^{11}\,\himsun}\right)^{1/3} \left(\frac{1+z}{4}\right)^{1/2} \kms. 
\label{eq:vc}
\end{eqnarray}
The parameter $\eta$ relates the halo circular velocity 
to the velocity width of the line. 
In principle, there is a broad distribution of velocity width 
depending on the local dynamics and geometry of 
the line emitting gas, resulting in a large scatter of $\eta$.  
\citet{Hae98} showed that, for DLAs, the median of 
the velocity width distribution is roughly 60\% of 
the virial velocity of the halo. 
Here we adopt $\eta=0.6$ following their work. 
Obviously this is an over-simplification of complex gas dynamics, 
but for now we are satisfied with this simple treatment. 
In the future we will compute the line profiles directly 
using the full velocity information in the simulations 
and study this issue in greater detail.

Using the \CII luminosity shown in Figure~\ref{fig:cii} 
and Equation~(\ref{eq:snu}), we obtain the \CII flux density 
for each halo at $z=3$ as shown in Figure~\ref{fig:snu_halo}. 
The distribution is roughly similar to that of \CII luminosity 
vs. halo mass, and can be characterized by 
\beq
\Snu = C_2 \left ( \frac{\Mhalo}{10^{12}\himsun}\right )^{2/3}, 
\label{eq:snu_scale}
\eeq
where $C_2 = 10^{-0.2} = 0.63$\,mJy (black {\it long-dashed} line, 
for the O3 run) and $10^{-1.2} = 0.063$\,mJy 
(red {\it short-dashed} line, for the Q5, D5, and G5 runs). 
The flux density follows the scaling $\Snu \propto \Mhalo^{2/3}$, 
because $\Lcii \propto \Mhalo$ and 
$\Dnu \propto v_c \propto \Mhalo^{1/3}$. 
We summarize these results in Figure~\ref{fig:snu_halo_summary}. 
This figure shows the lowest limiting halo mass one can probe 
for a given flux density limit. The limiting halo mass is
higher for the strong-feedback case than the no-feedback case, 
because the \CII flux density is lower in the strong-feedback case 
for a given halo mass due to smaller amount of CNM mass. 
The dispersion around the scaling relationships were roughly 
estimated by eye to be $\sim \pm 0.5$ dex 
from Figure~\ref{fig:snu_halo}. 

The cumulative flux density function is shown in Figure~\ref{fig:snu}. 
Here, we combine the results of the Q5, D5, and G5 runs 
to cover the entire range of flux density
as shown by the {\it black solid} line. 
The method for this interpolation is somewhat ad hoc, 
but the conclusion of this paper is not strongly affected 
by the details of this interpolation method. 
We simply require that the interpolated line 
(the `Combined' result) smoothly connects the result of 
different runs. 
Similarly to the \CII luminosity function, the runs 
with small box-sizes (Q5 and O3 runs) are limited at 
the bright-end owing to a lack of massive systems, and 
the run with large box-size is limited at the faint-end 
owing to a lack of low-mass systems caused by the limited resolution. 
The result of the O3 run is brighter than that of the Q5 run 
roughly by an order of magnitude.
Figure~\ref{fig:snu} suggests that the number density of galaxies 
at $z=3$ with flux density $\Snu > 0.1$\,mJy is about $2 \times 
10^{-2}\,\mpc^{-3}$ for the case of the `no-wind' (O3) run, 
and $3 \times 10^{-4}\,\mpc^{-3}$ for the `strong wind' run 
(the `Combined' result). 

Like Figure~\ref{fig:cii_lf}, we show the observed number density 
of LBGs with magnitudes $\Rab < 25.5$ at $z=3$ by \citet{Ade00} 
and \citet{Ade03} with the yellow shaded region. 
In addition, the two magenta {\it dot-long-dashed} curves 
are similarly obtained by using the same 
cumulative luminosity function of LBGs as in Figure~\ref{fig:cii_lf}. 
Inserting Equation~(\ref{eq:Rab_scale}) into 
Equation~(\ref{eq:snu_scale}) gives 
\beq
\log \Snu = -0.27\,(\Rab - C_3) + \log C_2 - 8. 
\eeq
For the case of the O3 run, this results in 
$\log \Snu = -0.27\,\Rab + 6.7$ (hereafter `O3-scaling'). 
We use this scaling and obtain the two magenta 
{\it dot-long-dashed} curves in Figure~\ref{fig:snu}:
the brighter one on the right is for the `O3-scaling', 
and the fainter one on the left is for the `Q5-scaling'.   
We also indicate the magnitudes $\Rab = 23.5$, 25.5, 30.0, and 36.0 
on this curve with filled triangles, which correspond to 
$\Snu = 2.3$, 0.7, 0.04, and $9.5\times 10^{-4}$\,mJy, respectively, 
for the `O3-scaling' case. 
The curve for the `O3-scaling' roughly agrees with the 
actual simulation result, but the `Q5-scaling' curve is 
in-between the actual O3 and the `Combined' results.  

Finally we look at the cumulative probability distribution function
as a function of \CII flux density in Figure~\ref{fig:snu_frac}. 
The striking fact is that the fraction of the sources with 
$\Snu > 0.1$\,mJy is very small at  $z=3$, i.e., less than 5\%. 
The fraction of LBGs brighter than $\Rab=30$ mag is also 
about the same level. 
This is due to the large number of faint sources 
in the CDM universe, and it is simply a reflection of 
a steeply increasing number of dark matter halos with 
decreasing halo mass. 
Using the observed luminosity function basically gives 
the same result.  The O3 run gives the most optimistic result 
in terms of the fraction of sources, which suggests that 
$\sim 30$\% of all sources have $\Snu > 0.01$\,mJy at $z=3$.


\section{Discussion \& Conclusions}
\label{sec:conclusion}

We have coupled state-of-the-art cosmological SPH simulations 
with an analytic model of a multiphase ISM in order to 
compute the \CII emission from galaxies at $z=3$. 
We find that, in a $\Lam$CDM universe, the majority of the sources 
are very faint with $\Snu < 0.1$\,mJy. 
This is presumably a generic prediction of the CDM model 
owing to an increasing number of low-mass dark matter halos 
with decreasing halo mass. 
If our model prediction on the faintness of the \CII sources 
is indeed correct, then it will be difficult for ALMA and SPICA 
to detect normal (i.e., not quasars, active AGNs, or strong 
submillimeter sources) high-redshift galaxies via \CII emission 
in large number, as the sensitivity limit of both telescopes are
expected to be $\Snu \sim 0.1 - 1$\,mJy. 
The recommended observing strategy is therefore to focus on 
very bright LBGs with $\Rab \ltsim 24$ mag that are pre-selected 
through optical imaging and spectroscopically known redshifts. 
Since there is a large body of spectroscopic data on 
LBGs at $z\gtsim 3$, it should not be a problem to come up with 
observing targets.  Our calculation shows that 
the brightest LBGs with $\Rab \sim 23.5$ mag could have 
flux densities $\Snu = 1-3$\,mJy depending on the strength of 
galactic wind feedback. 

Some caveats to note for the present work is that 
we have assumed that the physical conditions of neutral gas 
in DLAs and LBGs are the same, and that the dust-to-gas ratios 
of the gas are similar. 
If these assumptions are inappropriate for the LBGs at $z\sim 3$, 
then the above conclusions might not be entirely accurate. 
Our multiphase ISM model also assumed the SMC-type (silicates) 
dust composition which is less efficient for heating 
than other models such as the Galactic (carbonaceous) 
or the LMC-type models of dust grains. 
The true nature of dust in DLAs could be different from 
what we assumed here, and we expect a factor of two uncertainty 
in the dust-to-gas ratio. 
Note, however, that the metallicity of LBGs approaches solar, 
which is significantly higher than for DLAs. 
As a result, it is natural to expect that gas surrounding LBGs 
would have a dust-to-gas ratio higher than for the average DLA.  
Since the heating rate of the gas is proportional to 
dust-to-gas ratio, the \CII emission from LBGs is likely to be 
at the upper end of our calculation. 
In passing we note that star-forming DRG/BzK galaxies 
\citep{Dokkum04, Daddi04b} would also be good candidates 
for \CII observation as they are considered to have 
higher star formation rates and dust content than LBGs 
and hence more luminous in \CII luminosity. 
But their space density is lower by a factor of $\sim 10$ 
than that of LBGs. 

We have some confidence on the reliability of the properties
of bright galaxies in the simulation, because we know from 
our previous work \citep{NSHM, Night06, Finlator06} that 
the simulations can reproduce the observed properties of LBGs 
at $z=3-6$ relatively well.  
However the neutral gas is not well resolved 
near the resolution limit of each simulation, 
and this introduces some uncertainty for low-mass galaxies 
(except for the Q5 run in which low-mass galaxies are well modeled).
For this reason our scaling laws were mainly determined 
using the data for the massive galaxies in the simulation. 
There is also an intrinsic scatter in the CNM mass 
as a function of halo mass around the scaling relation 
(Figure~\ref{fig:nhmass}).  
As a result, we expect 30\% uncertainty in our predictions 
of the CNM mass fractions for high-mass halos, 
and somewhat larger uncertainties for lower mass halos.  

Even if the observation of normal high-redshift galaxies is difficult, 
the reward of \CII detection from such galaxies would be 
quite significant, because such measurements would directly 
constrain the amount of neutral gas and the properties of 
ambient ISM in high-redshift galaxies which is otherwise 
difficult to do. 
The observations of DLAs \citep[e.g.][]{Pro05, Wolfe05} have 
given us tremendous insights on the properties of neutral gas 
in high-redshift galaxies already, but \CII measurements will 
provide complementary information to further constrain 
theoretical/numerical models of galaxy formation. 
The combination of interferometric maps and the velocity widths 
of the \CII emission could in principle 
tell us about the physical sizes of DLAs and the mass of hosting 
dark matter halos, as well as the physical properties of the 
gas such as the heating rate. 

This paper is just a first step towards such a goal, 
and a number of improvements in the analysis are needed 
in the future. For example, in the current simulations, 
an optically thin approximation was assumed throughout 
when solving for ionization equilibrium, and a simple 
radiative transfer calculation assuming a disk geometry 
was used to approximate the effect of radiative transfer 
from local star-forming regions.  
As computing power increases, a more accurate treatment of 
radiative transfer \citep[e.g.,][]{Razoumov05} and 
ISM physics on small-scales will become possible 
using direct information from simulations, 
such as star formation rates and metallicity of the gas.


\acknowledgments 

We acknowledge the significant contribution of Volker Springel to the 
simulations used in this work, and we thank for his useful comments
on the manuscript.  We are also grateful to Alexei Kritsuk for useful 
suggestions and discussions.  This work
was supported in part by NSF grants ACI 96-19019, AST 00-71019, AST
02-06299, and AST 03-07690, and NASA ATP grants NAG5-12140,
NAG5-13292, and NAG5-13381.  The simulations were performed at the
Center for Parallel Astrophysical Computing at the Harvard-Smithsonian
Center for Astrophysics.


\appendix

\section{Appendix: Details of the multi-phase ISM model}

\begin{figure}
\epsscale{1.0}
\plotone{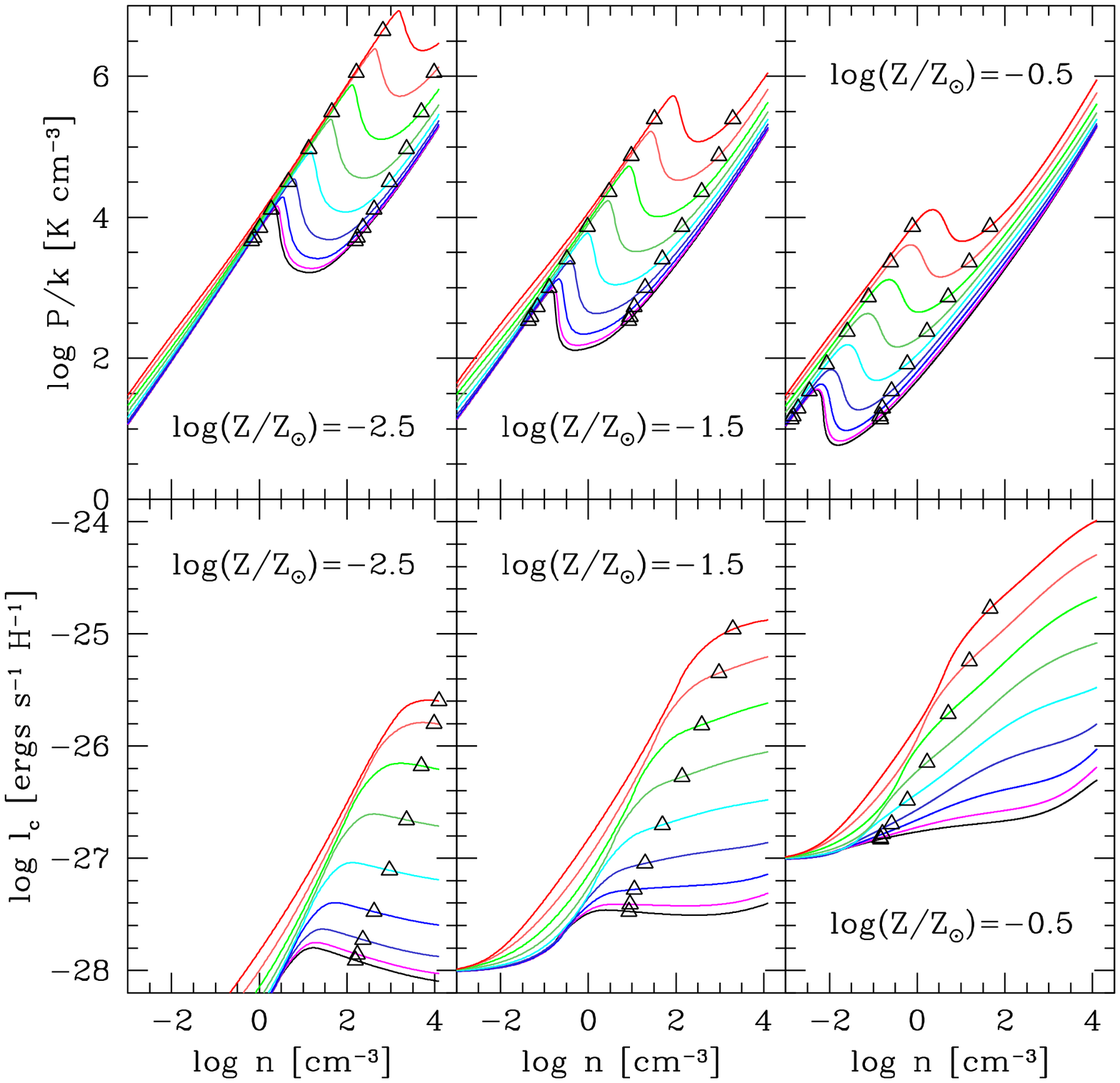}
\caption{{\it Top panels:} Phase diagram of pressure vs. density
for different projected SFR of $\log \psfr [\Msun\, \yr^{-1} \kpc^{-2}] 
= -4.0, -3.5, -3.0, -2.5, -2.0, -1.5, -1.0. -0.5$, and $0.0$, from bottom 
(black) to top (red).  
The open triangles indicate the CNM and WNM densities at the 
geometric mean pressure $\Pgeo = \sqrt{\Pmax \Pmin}$. 
{\it Bottom panels:} \CII luminosity per H atom, $\lc$, as computed 
by the model of \citet[][see Appendix A]{Wolfe03a}. 
See text for the qualitative trends of the curves shown in this figure. 
}
\label{fig:phase}
\end{figure}

\begin{figure}
\epsscale{1.0}
\plotone{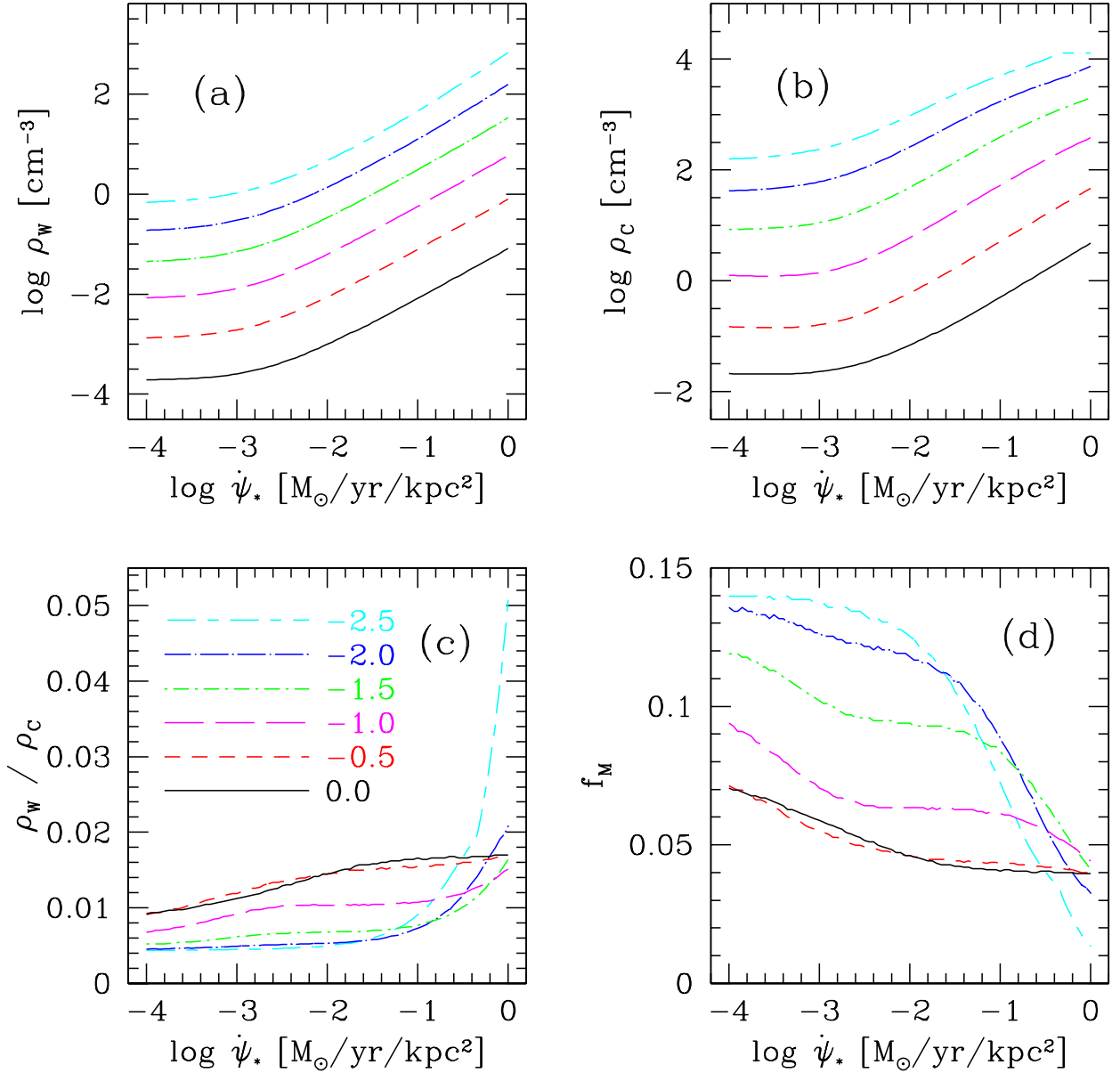}
\caption{Following quantities are shown as a function of projected 
SFR per unit physical area ($\psfr$ [$\Msun\, \yr^{-1} \kpc^{-2}$]): 
WNM density ($\log \rhow$, {\it panel [a]}), 
CNM density ($\log \rhoc$, {\it panel [b]}),
the ratio of the two ($\rhow / \rhoc$, {\it panel [c]}), and 
the mass fraction of CNM ($f_M$, {\it panel [d]}) as computed by 
Equation~(\ref{eq:fm2}) assuming a constant value of 
$f_V = 7\times 10^{-4}$ (see text). 
Different line types are for different metallicity as indicated in
the legend of {\it panel (c)}: 
$\log (Z/\Zsun) = -2.5$ ({\it long-dash short-dash}), $-2.0$ 
({\it long dashed-dot line}), $-1.5$ ({\it short dash-dot}), 
$-1.0$ ({\it long dashed}), $-0.5$ ({\it short dashed}), 
and $0.0$ ({\it solid}).
}
\label{fig:fm}
\end{figure}

\begin{figure}
\epsscale{1.0}
\plotone{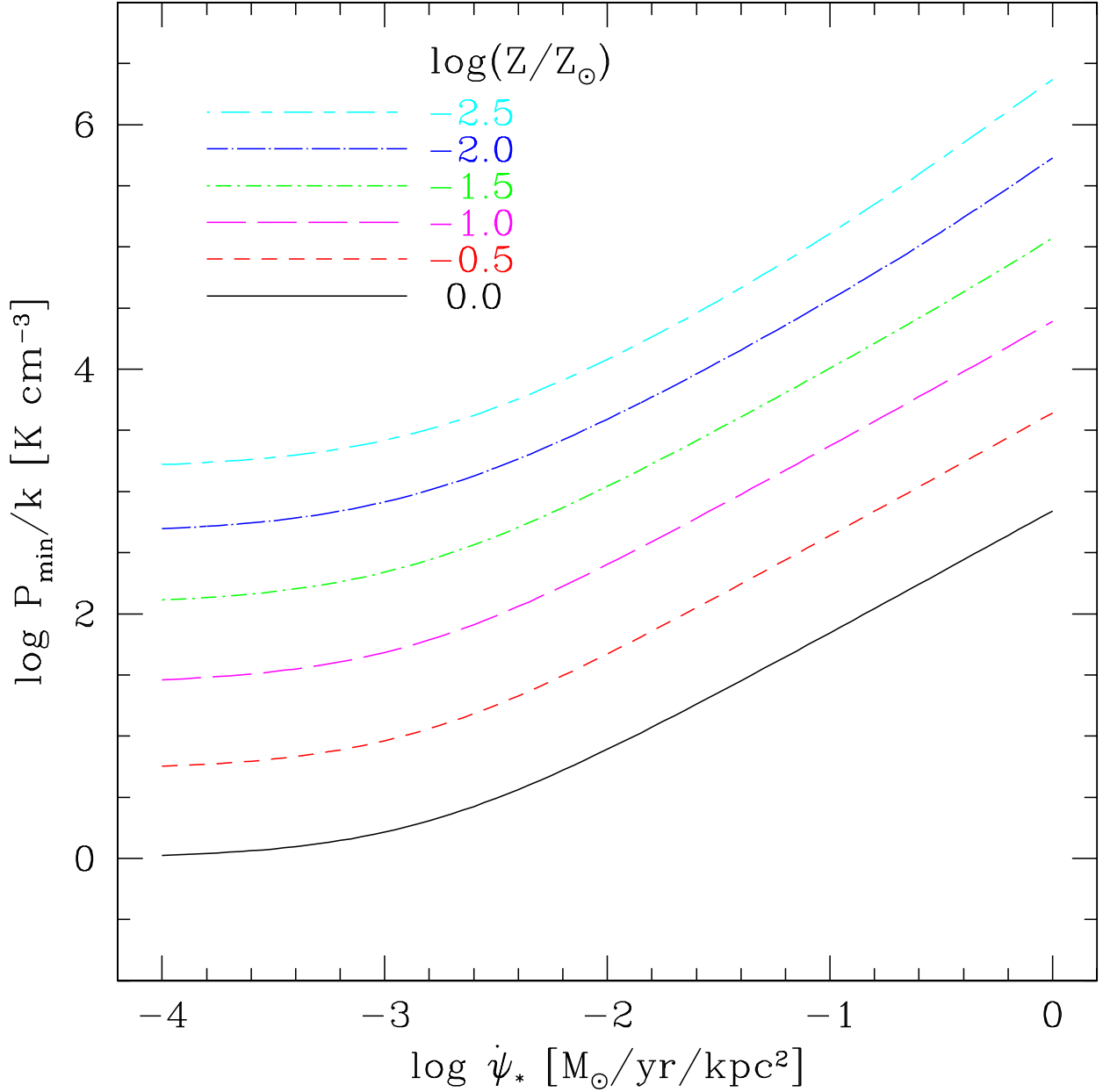}
\caption{Projected SFR $\psfr$ vs. minimum pressure $\Pmin$ described
in Section~\ref{sec:phase} and \ref{sec:CNMfrac}.  Different line 
types are for different metallicities as indicated in the legend. 
The simulated gas is required to satisfy $P>\Pmin$ and 
$\rho_0 > \rhow$ in order to host the CNM gas. 
}
\label{fig:pmin}
\end{figure}

\begin{figure}
\epsscale{1.0}
\plotone{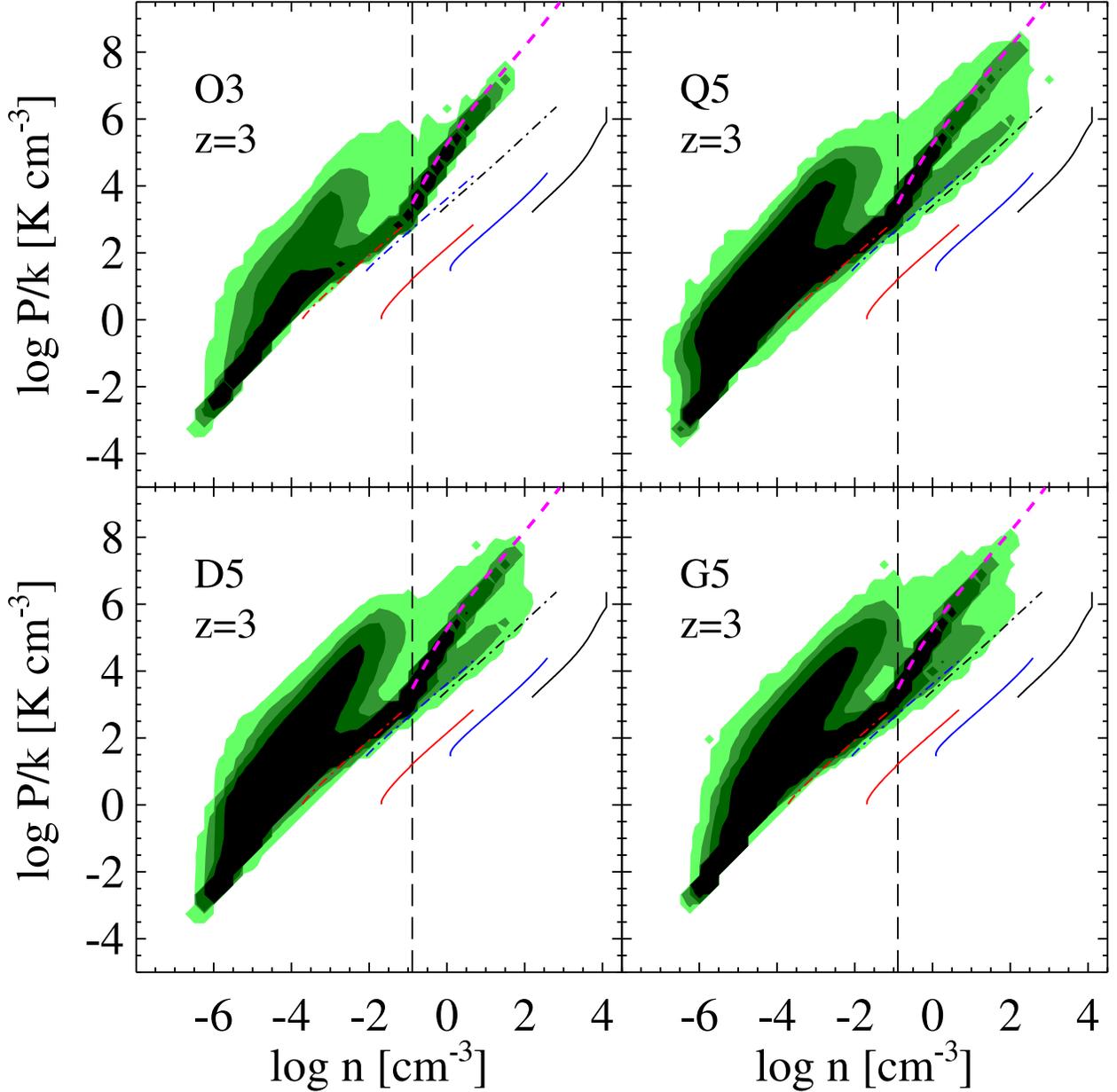}
\caption{Density vs. pressure of all cosmic gas in our O3, Q5, D5, 
and G5 simulations at $z=3$. The vertical dashed line indicates the 
threshold density $n_{\rm th} = 0.13$\,cm$^{-3}$ for star formation 
to take place in the simulation, and the gas to the right-ward of 
this threshold density is treated as ambient gas in star-forming 
regions in which the CNM and WNM are hosted. 
The 3 sets of 2 lines indicate the $\Pmin$ for metallicities of
$\log (Z/\Zsun) = -2.5$ (black), $-1.0$ (blue), and $0.0$ (red) as 
functions of WNM ({\it dot-dashed lines}) and CNM ({\it solid lines}) 
density at $\Pgeo$ from upper right to bottom left, respectively. 
The lines for the WNM densities are shown because the criteria 
that we impose for the simulated gas to qualify for hosting CNM 
are $P>\Pmin$ and $\rho_0 > \rhow$ (see Equations~[\ref{eq:fv}] and 
[\ref{eq:fm2}]).  
The magenta dashed line is the analytic fit to the effective equation 
of state adopted in the simulation \citep{Robertson04}. The 4 contour 
levels are for (1, 10, 100, 1000) gas particles in each 2-dimensional 
bin of size $(\Delta\log n, \Delta\log P/k)=(0.25,0.29)$ from low to 
high level.
}
\label{fig:pn}
\end{figure}

\begin{figure}
\epsscale{1.0}
\plotone{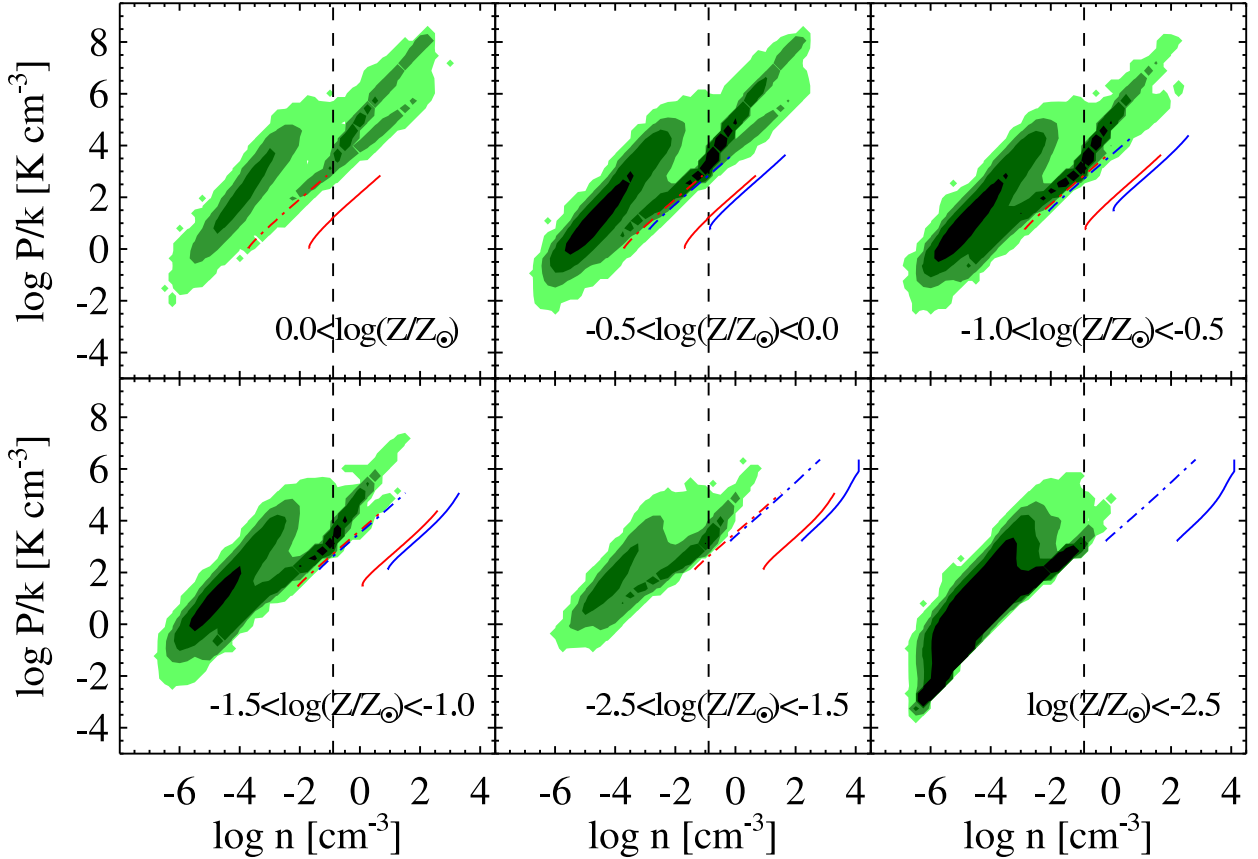}
\vspace{0.5cm}
\caption{Density vs. pressure of the gas in the Q5 run at $z=3$
for different metallicity ranges. 
The vertical dashed line is the same SF threshold density 
as shown in Figure~\ref{fig:pn}. The two sets of lines indicate 
the $\Pmin$ for the lower (blue) and higher (red) metallicity limit 
of each panel as functions of WNM ({\it dot-dashed lines}) and 
CNM ({\it solid lines}) density, similarly to those lines 
shown in Figure~\ref{fig:pn}. 
The lines for the WNM densities are shown 
because the criteria that we impose for the simulated gas 
to qualify for hosting CNM are $P>\Pmin$ and $\rho_0 > \rhow$ 
(see Equations~[\ref{eq:fv}] and [\ref{eq:fm2}]).  
The majority of the gas in the simulation that 
satisfy the pressure criteria $P>\Pmin$ for the CNM is high 
metallicity gas. 
The contour levels are the same as in Figure~\ref{fig:pn}.
}
\label{fig:pn_metal}
\end{figure}

\begin{figure}
\epsscale{1.0}
\plotone{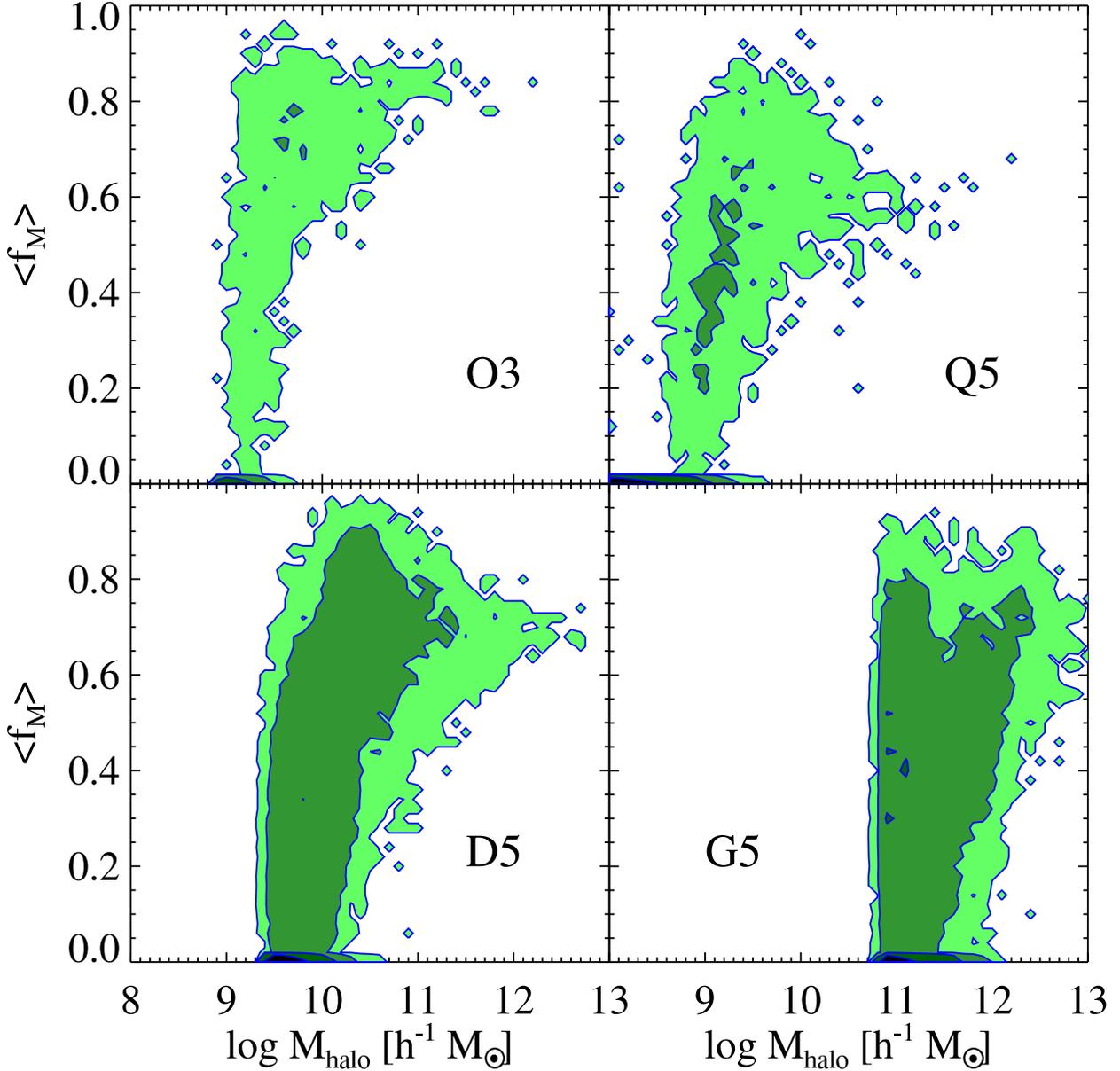}
\caption{Mean mass fraction of CNM of each halo, $\avg{f_M}$, as a 
function of halo mass at $z=3$. Each point on the figure represents 
the mean of 
$f_M$ over all grid cells that cover each halo. The contours
are used to avoid saturation in the scatter plot, and the four contour 
levels are for (1, 10, 100, 1000) data points in each two-dimensional 
bin of size $(\Delta\log \Mhalo, \Delta\avg{f_M})=(0.1,0.02)$ from low to 
high level. The two highest contour levels are not seen well
as there is a large pool of data points with $\avg{f_M}=0.0$, 
particularly for low mass halos in each run. 
The rapid decline of $\avg{f_M}$ values at $10 < \log \Mhalo < 12$ 
for the D5 and G5 run owes to lower resolution in simulations with 
large box-sizes, and we consider that the true decline occurs at 
around $\Mhalo \sim 10^{8.5}\himsun$ (see text).  }
\label{fig:fm_cont}
\end{figure}

\begin{figure}
\epsscale{1.0}
\plotone{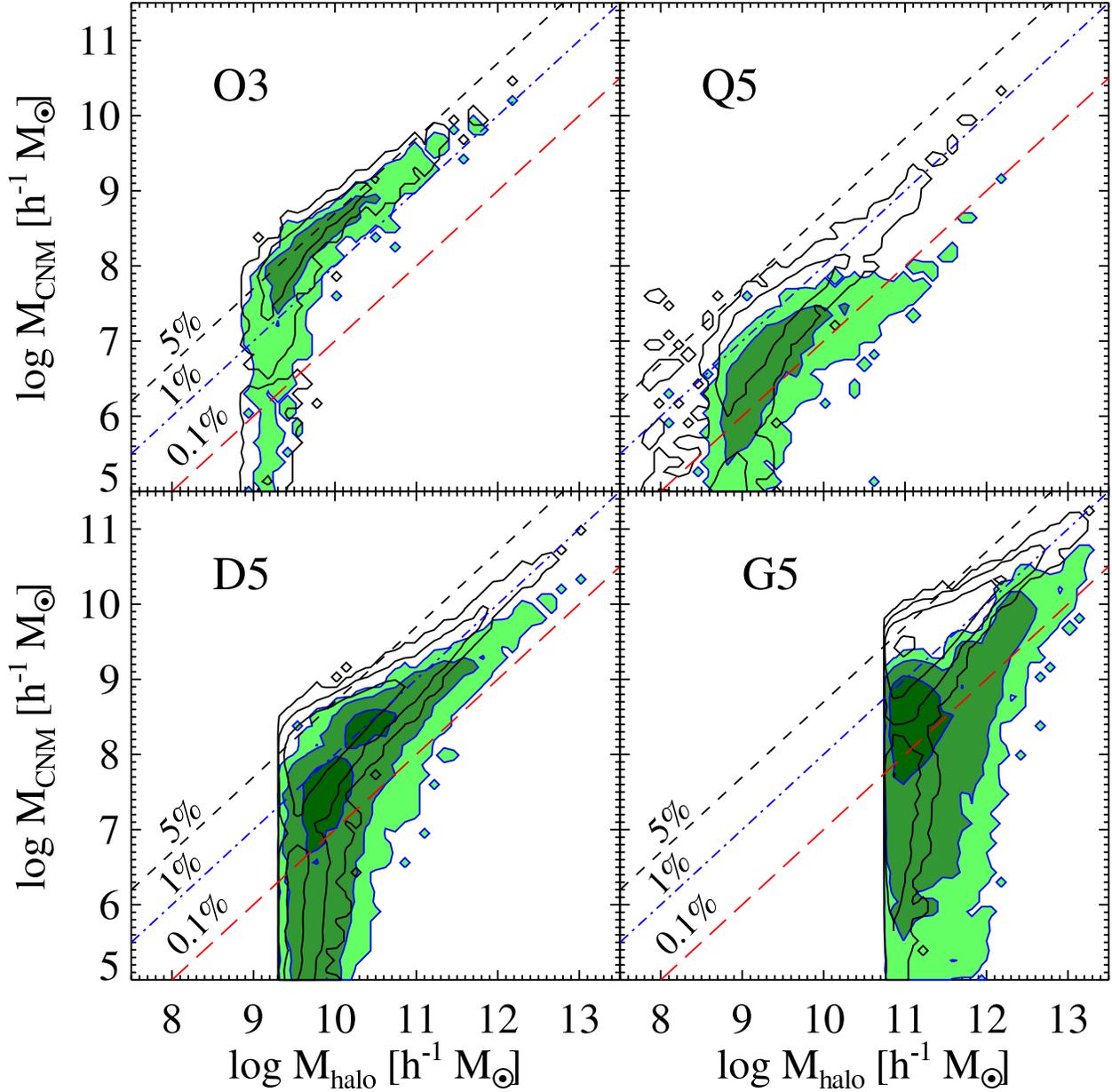}
\caption{Shaded contours show the CNM mass of each halo as a 
function of dark matter halo mass at $z=3$ in the simulations. 
The three contour levels are for (1, 10, 100) data points in each 
two dimensional bin of size $(\Delta\log \Mhalo, 
\Delta\log M_{\rm CNM})=(0.12,0.13)$ from low to high level.
The black contour lines without the shade is the total neutral
gas mass within each dark matter halo, i.e., the maximum amount of CNM 
that each halo could host.
}
\label{fig:nhmass}
\end{figure}

\begin{figure}
\epsscale{1.0}
\plotone{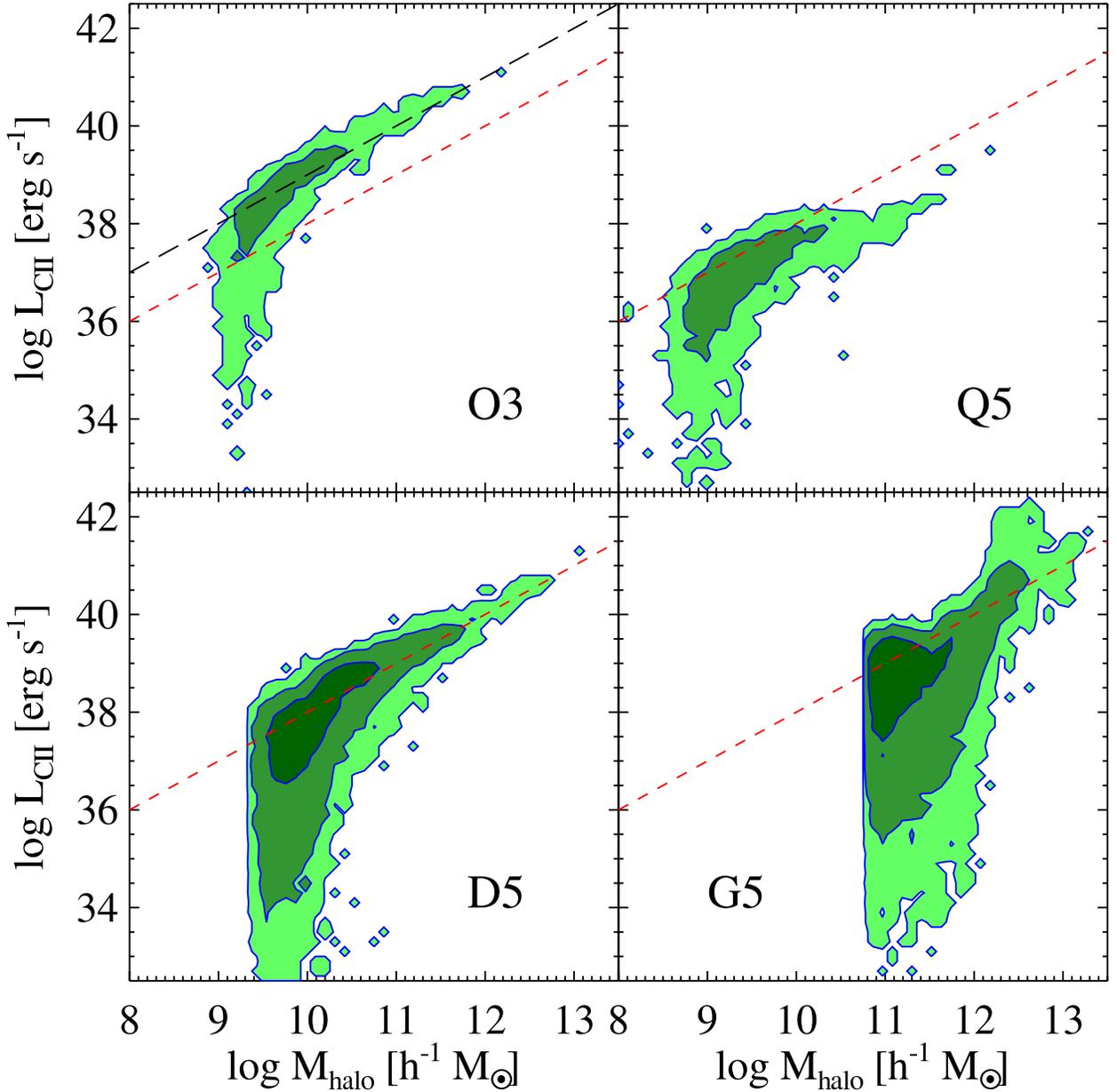}
\caption{\CII luminosity of each dark matter halo as a function of 
halo mass at $z=3$. Each point in this plot represents a single halo, but
we use contours to avoid saturation of points in the scatter plot. 
The 3 contour levels are for (1, 10, 100) data points in each 
2-dimensional bin of size $(\Delta\log \Mhalo, \Delta\log M_{\rm CNM}) = 
(0.11,0.20)$ from low to high level.
The {\it long-dashed} line in the top left panel and the {\it short-dashed} 
line in other panels show the relationship $\Lcii = C_1\,(\Mhalo/ 10^{12}\himsun)$, where $C_1 = 10^{41}$ and $10^{40}$\,erg\,s$^{-1}$, respectively. 
}
\label{fig:cii}
\end{figure}

\begin{figure}
\epsscale{1.0}
\plotone{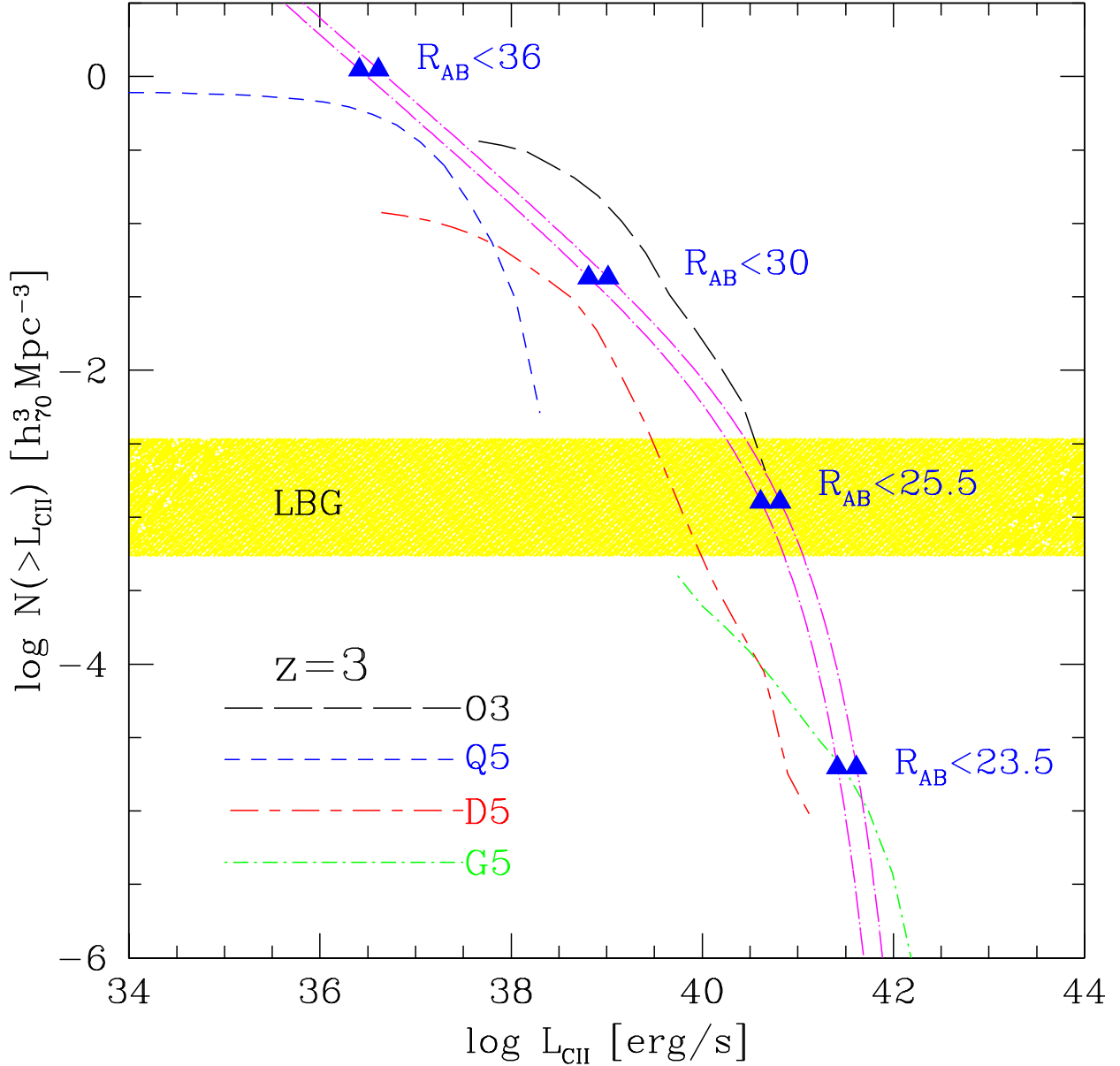}
\caption{Cumulative \CII luminosity function of simulations at $z=3$. 
The ordinate is in units of comoving $h_{70}^3 \mpc^{-3}$. 
The faint-end is truncated for the D5 and G5 run 
because of resolution limitations, and the bright-end of the O3 and Q5 runs 
is limited by cosmic variance owing to small simulation box-sizes.  
Note that the O3 run predicts a much higher \CII luminosity than the Q5 run 
because of less efficient galactic wind feedback which allows more neutral 
gas to remain within the dark matter halos and emit \CII line radiation.
The yellow shaded region indicates the observed number density
of LBGs brighter than $\Rab=25.5$ at $z=3$:   
$n_{\rm LBG} = 4\times 10^{-3}\,h^3\,\mpc^{-3}$ \citep{Ade00, Ade03}. 
See text for the details on the two magenta {\it dot-long-dashed} curves, 
which are derived from simple scaling laws between halo mass and $\Rab$ 
magnitude of LBGs and their observed luminosity function. 
}
\label{fig:cii_lf}
\end{figure}

\begin{figure}
\epsscale{1.0}
\plotone{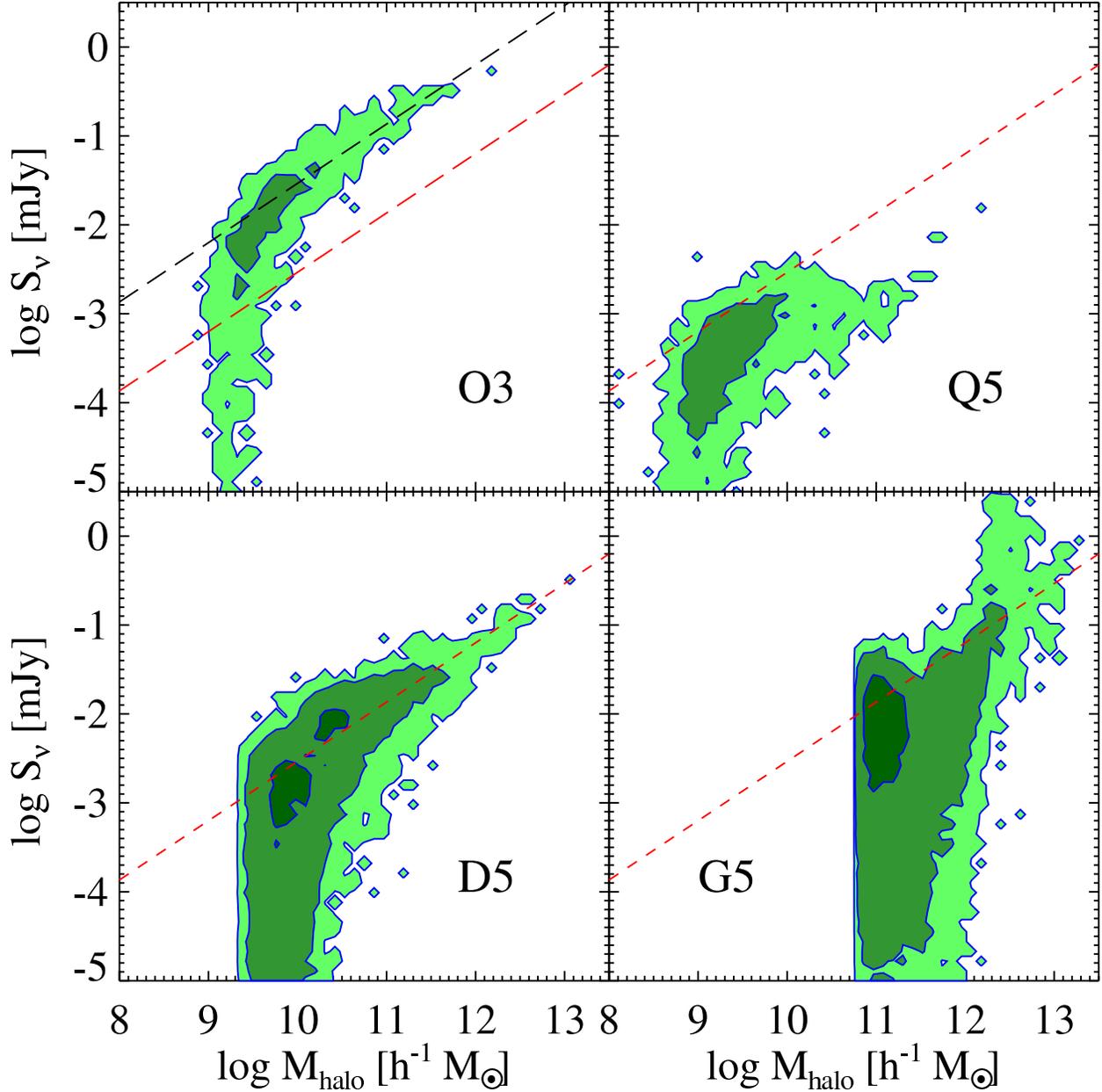}
\caption{\CII flux density of each dark matter halo as a function of 
halo mass at $z=3$. The 3 contour levels are for (1, 10, 100) data points 
in each 2-dimensional bin of size $(\Delta\log \Mhalo, \Delta\log S_\nu) 
= (0.11,0.13)$ from low to high.
The {\it long-dashed} line in the top left panel and the {\it short-dashed} 
line in other panels show the relationship $\log S_\nu = \frac{2}{3} 
(\log\Mhalo - 12) + C_2$, where $C_2 = -0.2$ and $-1.2$, respectively. 
}
\label{fig:snu_halo}
\end{figure}

\begin{figure}
\epsscale{1.0}
\plotone{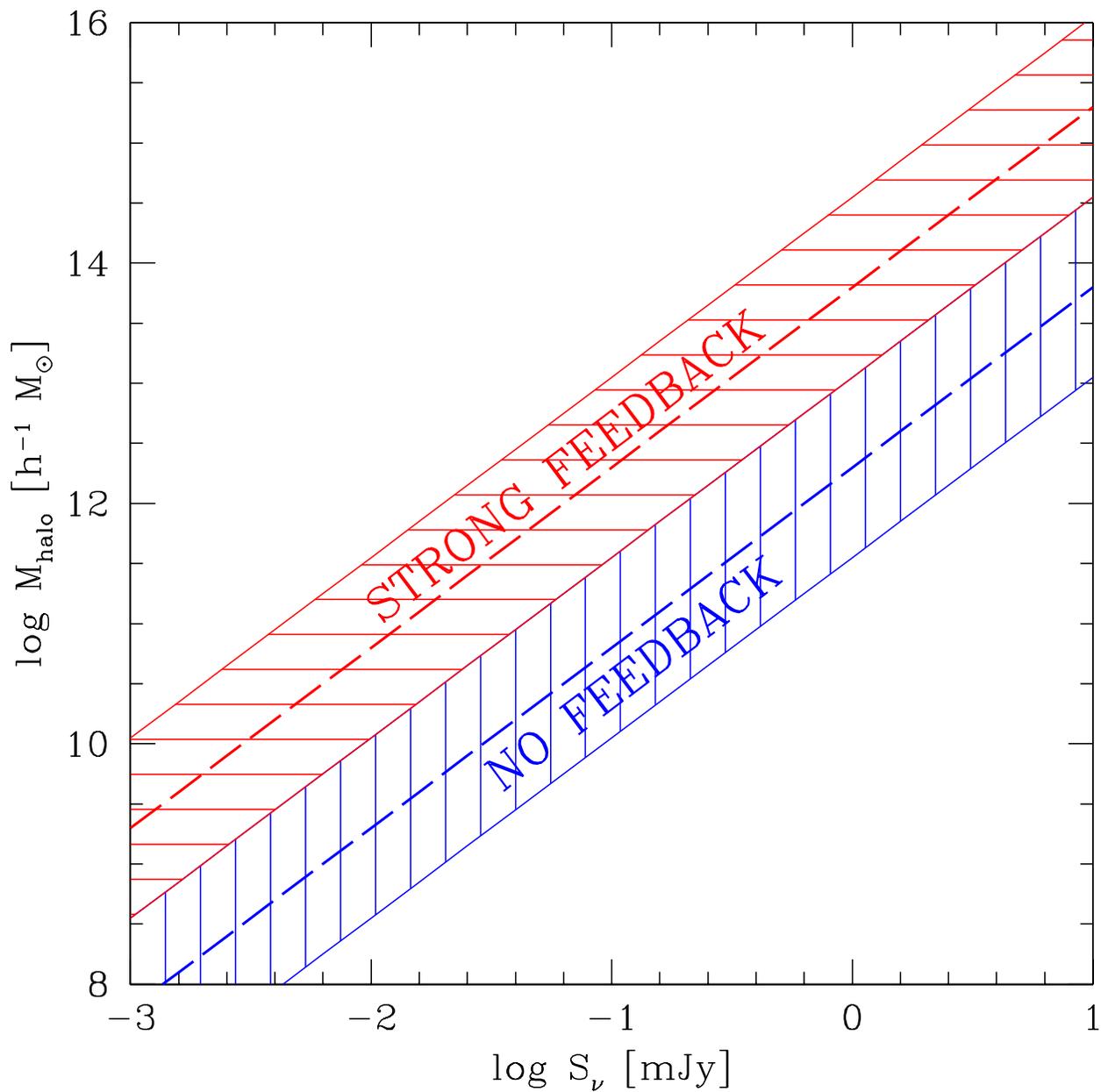}
\caption{
This figure shows the lowest limiting halo mass one can probe 
for a given flux density limit, summarizing Figure~\ref{fig:snu_halo}. 
The dashed lines are the same scalings shown in 
Figure~\ref{fig:snu_halo}, and the shaded region shows the 
dispersion of $\pm 0.5$ dex at a given flux density 
around the scaling relation. 
}
\label{fig:snu_halo_summary}
\end{figure}

\begin{figure}
\epsscale{1.0}
\plotone{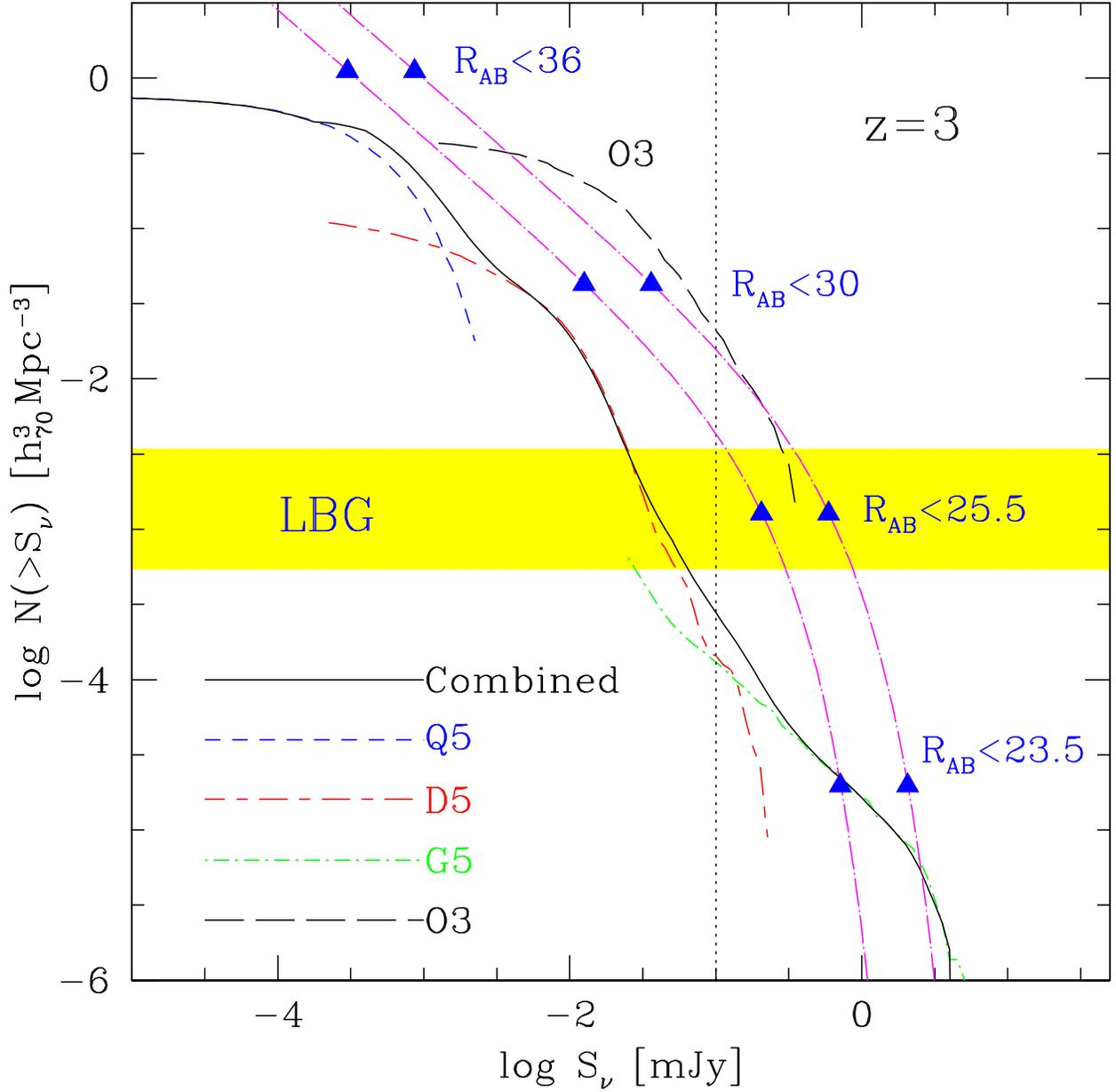}
\caption{Cumulative \CII flux density functions for O3, Q5, D5, G5 runs 
and the combined result of the latter 3 results.  
The ordinate is in units of comoving $h_{70}^3 \mpc^{-3}$. 
The rough detection limit of $\Snu = 0.1$\,mJy for ALMA and SPICA 
is indicated by the vertical dotted line. The difference between 
the results of the O3 and Q5 runs owes to the difference in the 
strength of galactic wind feedback. The yellow shaded region 
indicates the observed number density of LBGs brighter than 
$\Rab=25.5$ at $z=3$:   
$n_{\rm LBG} = 4\times 10^{-3}\,h^3\,\mpc^{-3}$ \citep{Ade00, Ade03}.
See text for the details on the two magenta {\it dot-long-dashed} curves, 
which are derived from simple scaling laws between halo mass and $\Rab$ 
magnitude of LBGs and their observed luminosity function. 
}
\label{fig:snu}
\end{figure}

\begin{figure}
\epsscale{1.0}
\plotone{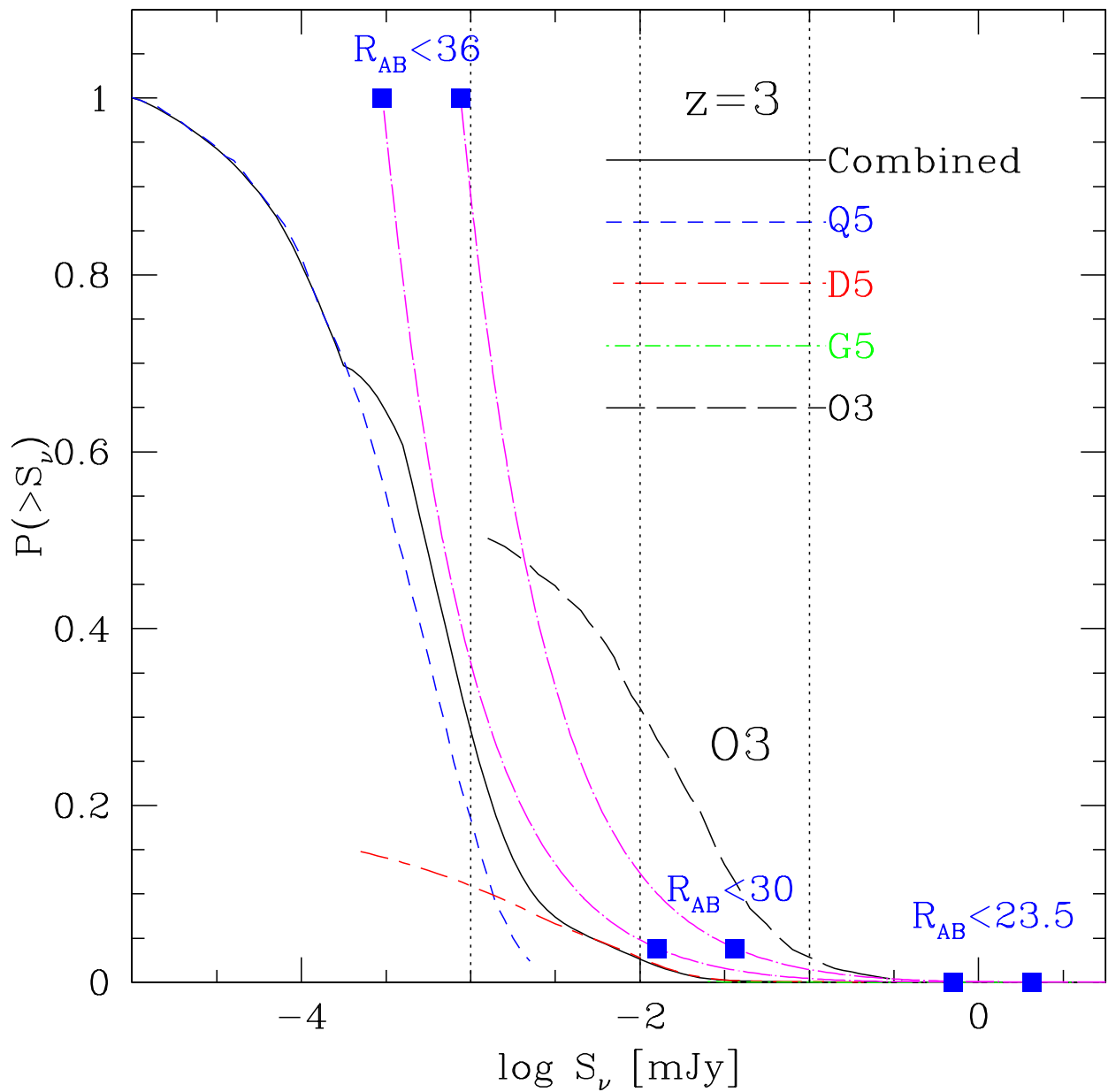}
\caption{Cumulative probability distribution of \CII sources as a function of
flux density $\Snu$ for the same models shown in Figure~\ref{fig:snu}. 
It is seen that the majority of the sources are faint objects with 
$\Snu < 0.1$\,mJy.  This suggests that one has to aim at very bright 
LBGs in order to have a detection even with ALMA and SPICA. 
}
\label{fig:snu_frac}
\end{figure}

Here, we briefly summarize the multi-phase ISM model developed by 
\citet{Wolfe03a}. The model is used for the calculation of 
Figure~\ref{fig:phase} in this paper.  

First, in order to relate the projected SFR per unit physical area, 
$\psfr$, with the incident FUV radiation field $J_\nu$, 
\citet{Wolfe03a} assumed a plane parallel disk geometry 
with half-width $h$ and radius $R$, in which the gas, dust, 
and stars are uniformly distributed. A disk is a reasonable 
approximation for dissipatively collapsing gas inside a dark matter
halo, and it is also motivated by the observations of DLAs
at $z\sim 3$ \citep[e.g.][]{Pro97, Pro98}. In the cosmological 
simulations used in this paper, we still lack the numerical 
resolution (and/or necessary physics) to properly account for 
the formation of disk galaxies in correct number, which is known 
as the `angular momentum problem' 
\citep[][and references therein]{Robertson04}. By assuming a disk
geometry for the ISM model, we are implicitly assuming that the 
simulated galaxies also have the same geometry, which is a reasonable
assumption. 

A radiative transfer calculation under this geometry gives 
\beq
J_\nu = \frac{1}{2}\left(\frac{\Sigma_\nu}{4\pi}\right)
\left [ 1 + \ln \left(\frac{R}{h}\right ) - k_\nu R 
\right ] + O(k_\nu R)^2 ...
\eeq
when $k_\nu h \ll k_\nu R \ll 1$. The quantity $\Sigma_\nu$ is the 
luminosity per unit area on the surface of the disk, and  $k_\nu$ is 
the absorption opacity of dust at frequency $\nu$. This equation 
shows that $J_\nu$ depends on the aspect-ratio $R/h$ only weakly. 
The relation $\Sigma_\nu = 8.4 \times 10^{-16}\, (\psfr\, /\, \Msun\, \yr^{-1} \kpc^{-2})$\, [erg\ cm$^{-2}$ s$^{-1}$ Hz$^{-1}$] is adopted from 
\citet{Madau00}. The effect of the UV background radiation is included
in $\Sigma_\nu$, but its effect is negligible compared to that owing to 
local star formation. 

Once the relationship between $J_\nu$ and $\psfr$ is obtained, 
we then compute the total heating rate $\Gamma$ for a given value 
of $\psfr$ as follows:
\beq
\Gamma = \Gamma_d + \Gamma_{\rm CR} + \Gamma_{\rm XR} + \Gamma_{\rm C_I}, 
\eeq
where the terms on the right-hand-side (RHS) are the heating rates owing to 
the grain photoelectric effect, cosmic rays, X-rays, and photoionization 
of C\,{\sc i} by the FUV radiation field $J_\nu$. 
In particular, the grain photoelectric heating rate per H atom,  
$\Gamma_d$, is related to $\psfr$ as 
\beq
\Gamma_d \propto \kappa\, \epsilon\, J_\nu \ [{\rm erg\ s^{-1} H^{-1}}]
\quad \propto \ \Sigma_\nu \quad \propto \ \psfr, 
\eeq
where $\kappa$ is the dust-to-gas ratio and 
$\epsilon$ is the fraction of FUV radiation 
absorbed by grains and converted to gas heating 
(i.e., heating efficiency). 
The $\Gamma_d$ was computed by adopting the \citet{Weingartner01}
expression for photoelectric efficiency in the case of pure silicates, 
blackbody FUV radiation, and extinction $R_V=3.1$. 
The heating rates $\Gamma_{\rm CR}$ and $\Gamma_{\rm XR}$ are also assumed
to be proportional to $\psfr$. 

The above heating rate has to be balanced with the total 
cooling rate $\Lam$ which includes the following terms:
\beq
\Lam = \Lam_{\rm FS} + \Lam_{\rm MS} + \Lam_{\rm Ly\alpha} + \Lam_{\rm GR}, 
\eeq
where the first term on the RHS is the cooling rate owing to 
fine-structure lines of \CII 158\,$\mu$m (which dominates at $T<300$\,K) 
and [O\,{\sc i}] 63\,$\mu$m (which is comparable to \CII at $T>300$\,K), 
i.e., $\Lam_{\rm FS} = \Lam_{\rm C_{II}} + \Lam_{\rm O_I}$.  
The second term on the RHS owes to metastable excitations of C$^+$, 
O$^0$, Si$^+$, and S$^+$, which becomes important at $T>3000$\,K.
The third and fourth term are the cooling by Ly$\alpha$ and grain
recombination which become important at high temperatures. Cooling owing 
to transitions in the neutral species C$^0$, Fe$^0$, Mg$^0$, and Si$^0$ 
are not included as their contribution to $\Lam$ is negligible. 

We then let 
\beq
\Gamma = n\,\Lam
\eeq
and solve the thermal and ionization equilibrium for given 
gas density $n$, dust-to-gas ratio, metallicity, and projected SFR. 
This leads to the two-phase gas structure as shown in the pressure 
curves in Figure~\ref{fig:phase}.
Once the thermal balance is achieved and CNM/WNM densities are
identified, the \CII emission per H atom can be obtained as
\beq
\lc = n\Lam_{\rm C_{II}} + \lc ^{\rm CMB},    
\eeq 
where the latter term on the RHS is the spontaneous energy emission 
rate in the limit of CMB excitation. The quantity $\lc$ is shown 
in the bottom panels of Figure~\ref{fig:phase} as a function of 
density.




\begin{deluxetable}{ccccccc}
\tablecolumns{2}  
\tablewidth{0pc}  
\tablecaption{Simulation Parameters}
\tablehead{
\colhead{Run} & \colhead{Box-size} & \colhead{$N_{\rm p}$} & \colhead{$m_{\rm DM}$} & \colhead{$m_{\rm gas}$} & \colhead{$\epsilon$} & \colhead{wind}
}
\startdata
O3  & 10.00 & $2\times 144^3$ &  $2.42\times 10^7$ & $3.72\times 10^6$ &2.78 & none \cr
P3  & 10.00 & $2\times 144^3$ &  $2.42\times 10^7$ & $3.72\times 10^6$ &2.78 & weak \cr
Q3  & 10.00 & $2\times 144^3$ &  $2.42\times 10^7$ & $3.72\times 10^6$ &2.78 & strong \cr
Q5  & 10.00 & $2\times 324^3$ &  $2.12\times 10^6$ & $3.26\times 10^5$ &1.23 & strong \cr
\hline                                                                  
D5  & 33.75 & $2\times 324^3$ &  $8.15\times 10^7$ & $1.26\times 10^7$ &4.17 & strong \cr
\hline                                                                  
G5  & 100.0 & $2\times 324^3$ &  $2.12\times 10^9$ & $3.26\times 10^8$ &8.00 & strong \cr
\enddata
\tablecomments{Simulations employed in this study.
The box-size is given in units of $\himpc$, ${N_{\rm p}}$ is the
particle number of dark matter and gas (hence $\times\, 2$), $m_{\rm
DM}$ and $m_{\rm gas}$ are the masses of dark matter and gas
particles in units of $\himsun$, respectively, $\epsilon$ is the
comoving gravitational softening length in units of $\hikpc$. }
\label{table:simulation}
\end{deluxetable}

\begin{deluxetable}{cccccc}  
\tablecolumns{2}  
\tablewidth{0pc}  
\tablecaption{Notation of Variables}
\tablehead{
\colhead{Variable} & \colhead{Definition} 
}
\startdata
$\rho_C  $, $\rho_W$ & density of CNM and WNM \\
$\rho_0$ & mean density of the total gas \\
$V_C$, $V_W$ & volume occupied by CNM and WNM \\
$V_0$ & volume of the star-forming region under consideration \\
$f_M$ & mass fraction of CNM \\
$f_V$ & volume fraction of CNM \\
$f_A$ & area covering fraction of CNM clouds \\
$\ncl$ & number density of CNM clouds \\
$R$ & characteristic radius of the spherical CNM cloud \\
$L$ & size of the star-forming region \\
\enddata
\label{table:notation}
\end{deluxetable}  


\end{document}